\documentclass[twocolumn,useAMS,usenatbib]{mnras}

\usepackage{amsmath, amssymb, ulem} 
\usepackage{multirow} 
\usepackage{pdflscape} 
\usepackage{graphicx}
\usepackage{color}
\usepackage[usenames,dvipsnames]{xcolor}

\newcommand\blfootnote[1]{%
  \begingroup
  \renewcommand\thefootnote{}\footnote{#1}%
  \addtocounter{footnote}{-1}%
  \endgroup
}

\newcommand{\redmapper}{redMaPPer} 

\newcommand{\red}[1]{#1}

\newcommand{\avg}[1]{\langle {#1} \rangle}
\newcommand{\Var}{\mbox{Var}}
\newcommand{\Mpc}{\mbox{Mpc}}
\newcommand{\msun}{M_{\odot}}
\newcommand{\lkhd}{{\cal{L}}}
\newcommand{\pcen}{p_{\rm cen}}
\newcommand{\Pcen}{P_{\rm cen}}

\newcommand{\hMsun}{h^{-1}\ \msun}

\newcommand{\scatname}{ESS}
\newcommand{\regauss}{re-Gaussianization}
\newcommand{\dsig}{$\Delta\Sigma$}
\newcommand{\photoz}{photo-$z$}
\newcommand{\pofz}{$p(z)$}
\newcommand{\xshear}{\texttt{xshear}}

\defcitealias{rmI}{RM I}
\defcitealias{rozoetal15b}{RM IV}

\title[The \redmapper{} Mass--Richness Relation]{Weak Lensing Measurement of the Mass--Richness Relation of SDSS \redmapper\ Clusters}
\author[Simet et al.]{Melanie Simet$^1$\footnote{Now at: Department of Physics and Astronomy, University of California, Riverside, CA 92507}, Tom McClintock$^2$, Rachel
  Mandelbaum$^1$, Eduardo Rozo$^2$, \newauthor Eli Rykoff$^3$, Erin
  Sheldon$^4$, Risa H. Wechsler $^{3,5}$\\
$^1$McWilliams Center for Cosmology, Carnegie Mellon University, 5000 Forbes Ave, Pittsburgh PA 15213, USA\\
$^2$Department of Physics, University of Arizona, 1118 E. Fourth St., Tucson, AZ 85721, USA\\
$^3$Kavli Institute for Particle Astrophysics and Cosmology, SLAC National Accelerator Laboratory, Menlo Park, CA 94305, USA\\
$^4$Brookhaven National Laboratory, Bldg 510, Upton, NY 11973, USA\\
$^5$Department of Physics, Stanford University, 452 Lomita Mall, Stanford, CA 94305-4085, USA}

\begin{document}

\maketitle

\begin{abstract}
We perform a measurement of the mass--richness relation of the \redmapper{} galaxy cluster catalogue using
weak lensing data from the Sloan Digital Sky Survey.   
We have carefully characterized a broad range of systematic uncertainties,
including shear calibration errors, \photoz\ biases, dilution by member galaxies, source obscuration,
magnification bias, incorrect assumptions about cluster mass profiles, cluster centering, halo triaxiality, and projection effects. 
We also compare measurements of the
lensing signal from two independently-produced shear and photometric redshift catalogues to characterize systematic errors in the lensing signal itself.  
Using a sample of 5,570 clusters from $0.1\le z\le 0.33$, the normalization of our power-law mass vs.\ $\lambda$ relation is 
$\log_{10}[M_{200m}/\hMsun]$ = $14.344 \pm 0.021$ (statistical) $\pm 0.023$ (systematic) at a richness $\lambda=40$, a 7 per
cent calibration uncertainty, with a power-law index of $1.33^{+0.09}_{-0.10}$ ($1\sigma$). 
The detailed systematics characterization in this work 
renders it the definitive weak lensing mass calibration for SDSS \redmapper\ clusters at this time.
\end{abstract}

\section{Introduction}

\blfootnote{$^{\star}$Now at: Department of Physics and Astronomy, University of California, Riverside, CA 92507}

The abundance of galaxy clusters is a powerful cosmological probe \citep{DarkEnrgTF}.  In this work, we 
measure the relationship between the weak lensing masses and the optical richness 
of galaxy clusters.  Weak lensing offers a unique key to 
understanding the masses of structures in the 
universe, due to its equal sensitivity to dark and baryonic matter \citep{2006glsw.conf....1S}.  
Galaxy clusters are a good target for weak lensing due to their high masses and thus large lensing distortions.  
Using weak lensing mass measurements, then, we can better understand the relationship between
cluster masses and other observables, which aids scientific goals such as 
measurements of cluster abundances for cosmology.

Weak lensing exploits the deflection of light rays and the resulting distortion of galaxy shapes by gravitational fields.  By 
measuring statistical changes in the shapes of more distant galaxies, we can detect the 
gravitational fields sourced by intervening matter and therefore probe the distribution and amount of
matter in the universe \citep{2006glsw.conf....1S}.  Weak lensing has been commonly 
used to characterize the matter in galaxies 
\citep[e.g.][]{1996ApJ...466..623B}, groups \citep[e.g.][]{2010ApJ...709...97L} and galaxy clusters 
\citep[e.g.][]{2004AJ....127.2544S,2007arXiv0709.1159J,2014MNRAS.439....2V,2015MNRAS.449..685H,okabeetal15,2016A&A...586A..43V}, as well as the large-scale distribution of matter through the 
technique known as cosmic shear
\citep[e.g.][]{2000MNRAS.318..625B,2013MNRAS.430.2200K, 2015arXiv150705598B}.   

Galaxy clusters, the target of this weak lensing study,  
represent the most massive gravitationally bound structures in the universe 
\citep[e.g.,][]{2011ARAA..49..409A}.  They consist of multiple galaxies in a large dark matter halo, usually with 
one large elliptical galaxy in the centre.  These galaxies are 
systematically different from other elliptical galaxies \citep[e.g.][]{2007MNRAS.379..867V}.  The dominant 
baryonic component is a reservoir of hot gas held by the potential well of the dark matter halo, but 
this gas is visible only in the X-ray (thermal Bremsstrahlung) and radio (through scattering of 
photons from the cosmic microwave background).  All of these properties can be used to construct 
cluster catalogues, based on characteristics such as distortions of
the observed spectrum of CMB photons along the line of sight to the clusters 
\citep{2015ApJS..216...27B}, extended X-ray emission \citep{1982ApJ...253..485P}, or simply an overdensity of 
optically-detected galaxies at the same redshift.  In this paper, we use the 
\redmapper{} optical cluster finder \citep[][henceforth RM I]{rmI}, described in Sec.~\ref{sec:rm}.  
Cluster catalogues typically have a ranking mechanism based on a mass proxy such as X-ray luminosity or the number of galaxies 
in the cluster.  We calibrate the relationship between the \redmapper{} mass proxy, the 
summed galaxy membership probability $\lambda$ (also known as the optical richness), and the masses derived from weak lensing measurements.

Previous work on the \redmapper\ catalogue mass--richness relation has included comparison to other cluster data sets and
weak lensing measurements from several different imaging surveys 
\citep{rozorykoff14, 2015MNRAS.454.2305S, 2015arXiv150606135M, 2015arXiv151008193D}.
Due to widely varying 
parameterization choices, we will discuss these results further in Section~\ref{conclusions}, 
including conversions necessary to compare amongst them and against our work.  The work presented here
is the largest sample for which such measurements have been made; we obtain high signal-to-noise
measurements of the lensing signal in several richness bins and two redshift bins, and also use what 
we believe to be the most complete model for the lensing signal and its systematics.  
Our results are consistent with previous measurements, but precisely calibrate the systematic uncertainty
associated with the weak lensing masses.

We discuss the background of our weak lensing procedure in Sec.~\ref{bg-wl}.  We then discuss the 
\redmapper{} algorithm and its application to SDSS DR8 data, particularly its richness 
estimator $\lambda$, in Sec.~\ref{sec:rm}.  The lensing shear catalogue is described in
Sec.~\ref{sec:lenscat}.  The mass model that we use is detailed in Sec.~\ref{modeling}, 
and we address a variety of sources of systematic error in Sec.~\ref{sec:sys}.
Our results from model fits are given in Sec.~\ref{results}.  We summarize and conclude in Sec.~\ref{conclusions}.  Throughout the paper, except where 
noted, we assume a flat $\Lambda$CDM cosmological model with $\Omega_m=0.3$, $\sigma_8$=0.8, and $h_{100}=1$.
Unless otherwise specified, all distances are physical distances (rather than comoving).

\section{Weak lensing background}\label{bg-wl}

The deflection of light by gravity affects the apparent shape, size, and number
density of galaxies.  These effects
can be used to measure the relationship between dark matter and visible matter, or more generally
to probe cosmological models \citep{2001PhR...340..291B, 2002PhRvD..65f3001H, 2002PhRvD..66h3515H, 
2003PhRvL..91d1301A, 2003ARAA..41..645R, 2006glsw.conf....1S,
2008ARNPS..58...99H, 2010RPPh...73h6901M}.
An overdensity of matter, such as a galaxy or galaxy cluster, will cause 
a slight tangential alignment of galaxies at more distant redshifts; an underdensity, such as a void, 
will cause a slight radial alignment.  In addition, the number density \citep[e.g.][]{broadhurstetal95} 
or distribution of quantities 
such as redshift \citep[e.g.][]{couponetal13} and size \citep[e.g.][]{2009PhRvL.103e1301S} will be altered, 
due to effects that shrink or enlarge the galaxy images 
on the sky.  Shape distortion is usually called the lensing shear, while the other effects come 
under the umbrella of lensing magnification.  Since weak lensing (as opposed to strong 
lensing) results in only slight differences in observed galaxy properties, it must be measured 
statistically.  

For galaxy cluster lensing, we are well within the thin-lens limit, where the line-of-sight extent 
of the lens is much smaller than distances between the observer and the lens, and between the
lens and the lensed galaxy.  
When considering lensing shear, we are sensitive to the surface mass density $\Sigma(R)$, where $R$ 
here is the projected separation on the sky from the lens, calculated as the angular diameter distance $D_A(z)$ times the angular separation between the centre of the 
lens and the lensed galaxy.  The tangential shear $\gamma_t(R)$ imposed by a lens at redshift
$z_l$ on a background source at redshift $z$ is given by~\citep{2006glsw.conf....1S}
\begin{equation}
\gamma_t(R) = \frac{\bar{\Sigma}(<R) - \bar{\Sigma}(R)}{\Sigma_{\mathrm{cr}}(z_l,z)},
\end{equation}
that is, the average surface mass density interior to the radius $R$ minus the average surface mass 
density in an annulus at radius $R$, modulated by a geometric factor known as the critical surface 
mass density:
\begin{equation}
\Sigma_{\mathrm{cr}}(z_l, z) = \frac{c^2}{4\pi G} \frac{D_A(z)}{D_A(z_l)D_A(z_l,z)}.
\end{equation}
In practice,
since we use lenses and sources at many different redshifts, we formulate the problem as an estimate
of $\Delta\Sigma$
\citep{McKayGGL2001},
\begin{equation}\label{deltasigma}
\Delta\Sigma \equiv \Sigma_{\mathrm{cr}}(z_l,z)\gamma_t(R) = \bar{\Sigma}(<R) - \bar{\Sigma}(R),
\end{equation}
which, being directly related to the projected mass profile around the lens,
is the same for identical lenses regardless of the lens and source redshifts. 

Since the lensing-induced shear is significantly smaller than the intrinsic scatter in galaxy 
shapes, we use a statistical estimator that sums over all background (lensed) galaxies $i$ found 
within an annulus centred at radius $R$ around the lens: 
\begin{equation}\label{DeltaSigmaEstimator}
\widehat{\Delta\Sigma}(R) = C(R) \frac{\sum_i w_{i} \Sigma_{\mathrm{cr}} (z_l,z_i)\gamma_{t,i}}{\sum_i w_{i}}.
\end{equation}
%
The optimal weights $w_i$ above include
both the per-object shape noise weighting 
$w_{i,\mathrm{shape}}$ and the critical surface mass density~\citep{SheldonGGL2004}, 
\begin{equation}\label{Weights}
w_i = \Sigma_{\mathrm{cr}}^{-2} w_{i,\mathrm{shape}},
\end{equation}
where $w_{i,\mathrm{shape}} = 1/(\sigma_{\rm int}^{2} + \sigma_i^2)$ includes both intrinsic shape noise and measurement 
uncertainty.
The per-object shape noise is a product of our shape measurement and is described in 
Section~\ref{sec:lenscat}.   The inclusion of the critical surface mass density optimally accounts
for the different lensing geometries of galaxies at different redshifts, though the effects of
photometric redshift error prevent us from achieving an optimal measurement \citep{2008MNRAS.386..781M}.
We explicitly cancel out factors of $\Sigma_{\mathrm{cr}}$ in the numerator of 
Equation~\eqref{DeltaSigmaEstimator} to highlight that the summand converges to 0 as $z_i$
approaches $z_l$ from above.

In practice, we compute the average signal for many lenses at once.  This increases our 
signal-to-noise ratio and also reduces biases due to individual characteristics of the lens (such as 
extra line-of-sight overdensities, lens substructure, and lens asphericity) by averaging those 
properties as well \citep[e.g.][]{2007ApJ...656...27J, 2009MNRAS.396..315C}. So, for a bin 
containing lenses $j$, we in fact have 
\begin{equation}\label{DeltaSigmaEstimatorStack}
\widehat{\Delta\Sigma}(R) = C(R) \frac{\sum_{i,j} w_{ij} \Sigma_{\mathrm{cr},ij}\gamma_{t,i}}{\sum_{i,j} w_{ij}}.
\end{equation}

We include a boost factor $C(R)$ to account for contamination of our background
galaxy sample by galaxies that are
physically associated with the cluster \citep{FischerGGL2000}.  Our background galaxy source 
sample is photometric, with the catalogue described in Section~\ref{sec:lenscat}.  Because of the scatter 
in the photometric redshift estimates, some galaxies located in the cluster will 
be scattered into the background sample; these galaxies have a shear expectation value of 0, unlike 
the lensed background galaxies, so they will dilute the measured lensing signal.  This is a 
particular problem for galaxy clusters, since they contain many galaxies that can be scattered this 
way.  
To correct for the dilution effect from cluster galaxies in the source population,
we compute the weighted number density of sources around random points 
(described in Section~\ref{sec:rm}) using the weights from Eq.~\eqref{Weights}.  For 
background galaxies $i$ around $N$ lenses $j$, and background galaxies $k$ around $N_{\mathrm{rand}}$
random points $l$:
\begin{equation}\label{boosteq}
 C(R) = \frac{N_{\mathrm{rand}}}{N}\frac{\sum_{i,j} w_{ij}} {\sum_{k,l} w_{kl}}.
\end{equation}
\red{This multiplicative correction is performed on a per-bin basis, and ranges from 1.2-1.4 
in the innermost radial bin used for fitting in the lower-redshift cluster bins or 
from 1.4-1.7 in the innermost radial bin used for fitting in the higher-redshift cluster bins.}
We note that the above correction factor can be explicitly demonstrated to be correct \citep{rozoetal11},
and is commonly used in statistical lensing measurements \citep[e.g.,][and associated
follow-up work]{SheldonGGL2004,mandelbaumetal05}.
Indeed, this is one of the great advantages of using photometric survey data and treating
weak lensing as a statistical measurement; \red{by using a stack of many objects for which the
  source density converges to a well-understood average across the survey, we can measure and correct for
this contamination very accurately without worrying about statistical fluctuations in the
populations behind individual clusters, and without having to use overly-conservative cuts that may
risk removing many lensed objects.}

To account for the survey mask, which otherwise might impose a small signal at large radii
  due to systematic errors that correlate with the survey boundaries, we subtract the 
$\Delta\Sigma$ estimate around random points from our real signal before the boost factor is applied
\citep[see, e.g.,][]{mandelbaumetal05,2013MNRAS.432.1544M}\red{, also on a per-bin basis}.

\section{Data}

In this work, we use data from the Sloan Digital Sky Survey \citep[SDSS;][]{2011AJ....142...72E, 2000AJ....120.1579Y}.  
Both our cluster and weak lensing catalogues derive from data release
8 \citep[SDSS DR8][]{2011ApJS..193...29A}.  

\subsection{The \redmapper\ Cluster Catalogue}\label{sec:rm}

\redmapper\ is a red-sequence photometric cluster finding algorithm \citep[][hereafter RM I]{rmI}, 
built around the optimized 
richness estimator developed in \citet{rozoetal09b} and \citet{rykoffetal12}. 
Briefly, \redmapper\ identifies galaxy clusters
as overdensities of red galaxies, and estimates the probability that each red galaxy is a cluster member following
a matched filter approach which models the galaxy distribution as the sum of a cluster and background component.
The cluster richness $\lambda$ is the sum-total of the membership probabilities of all the galaxies.  
The cluster radius $R_\lambda$ used for estimating the cluster richness is self-consistently computed
with the cluster richness, ensuring that richer clusters have larger cluster radii.  The radius $R_\lambda$
is selected to maximize the signal-to-noise of the richness measurements.
The \redmapper\ cluster 
richness $\lambda$ has been shown to be tightly correlated with cluster mass by comparing $\lambda$ to well-known
mass proxies such as X-ray gas mass and SZ decrements.  The original (v5.2) \redmapper\ algorithm was published in 
\citetalias{rmI}, to which the reader is referred for further details.  Here, we utilize the v5.10 version of the algorithm,
which introduced a variety of small improvements; we refer the reader to \citet[][hereafter RM IV]{rozoetal15b} for details.

The \redmapper\ algorithm was applied to SDSS DR8 data
\citep{2011ApJS..193...29A} in \citetalias{rozoetal15b}.  The
\redmapper\ catalogue is restricted to the $\sim 10,000\ \deg^2$
of contiguous imaging used by the Baryon Oscillation Spectroscopic Survey
\citep[BOSS;][]{2013AJ....145...10D}. Roughly 2/3 of the survey falls
in the Northern Galactic cap, and 1/3 in the South.  The sky was imaged in 5 bands ($ugriz$), and we have imposed a limiting
magnitude $i<21$, which is a conservative estimate for the full
footprint~\citep[see][]{2015arXiv150900870R}.  \redmapper\ utilizes the 5-band imaging
data to self-calibrate a model for red-sequence galaxies, and then applies this model to identify red galaxy overdensities and
to estimate the corresponding photometric redshift of the galaxy clusters.  SDSS \redmapper\ photometric redshifts are accurate
at the $0.005$ to $0.01$ level, depending on redshift. Here, we ignore the uncertainty associated with the cluster
photometric errors.  We have verified that randomly perturbing every cluster by its assigned photometric redshift
uncertainty impacts our conclusions at less than the 1\% level, demonstrating that the cluster photo-$z$ uncertainties
are indeed negligible for this analysis. 
  
The \redmapper\ algorithm explicitly assumes that the centre of each cluster's halo coincides with the location
of one of the brightest galaxies in the cluster, though not necessarily the brightest.  Indeed,
an important feature of \redmapper\ is that it does not simply choose a cluster centre, it also attempts
to estimate the probability $\pcen$ that each \redmapper\ cluster is properly centred.  The probability is estimated
based on each galaxy's luminosity, photometric redshift, and local galaxy density.  \redmapper\ relies on 
a Bayesian classification scheme with empirical, self-calibrated filters
for the distribution of central and satellite
galaxy properties, modified to take into account that every cluster contains one and only one central
galaxy.  For
details, we refer the reader to \citetalias{rmI}. 
Given the probability $\pcen$ for each cluster, the fraction of well-centred
clusters over the entire cluster catalogue is simply $\Pcen = N_{\rm clusters}^{-1}\sum \pcen$, where the sum
is over all clusters in the catalogue.   
As discussed in section~\ref{sec:miscentre}, the fraction $\Pcen$ is an important systematic parameter in our
analysis.  

Cluster random points were generated using the updated method of
\citet{2016arXiv160100621R}.  In brief, we first sample a cluster from the \redmapper\ cluster catalogue.  This gives us the richness
and redshift of our random point.  We then randomly select a right ascension and declination within the survey footprint.  If the survey is
not sufficiently deep to have detected a cluster at that location, we reject the cluster, while keeping track of the
number of times $N_R$ which each cluster is rejected.  If the survey is sufficiently deep to detect our randomly selected
cluster, then the cluster is added to the random catalogue at the randomly drawn position.  The procedure is iterated
until the random catalogue achieves the desired number of random points.  Let $N$ be the number of times
that a particular cluster appears in the random catalogue, and $N'$ by the number of times the cluster was rejected
as a random point.  We assign a weight $w=(N+N')/N$ to each instance of this cluster in the random catalogue.  This ensures
that the weighted random catalogue exactly traces the richness and redshift distribution of the parent cluster catalogue.  

To perform the weak lensing measurement, we bin the clusters in 4 richness bins and 2 redshift
bins, for a total of 8 bins.  The bins are detailed in Table~\ref{bintable}.  We note that the largest bin
has a very broad range of $\lambda$, but we checked alternate binning schemes and found no statistically
significant differences between our final results. 
Our fiducial binning scheme has roughly equal signal-to-noise for all richness and redshift bins.

\begin{table*}
	\centering
	\begin{tabular}{c c | c c | c | r}
		\hline
		& & & & & No. of \\
		$\lambda$ range & mean $\lambda$ & $z$ range & mean $z$ & mean $\Pcen$ & clusters \\
		\hline
		$[20,30)$ & 24.1 & $[0.1,0.2)$ & 0.153 & 0.87 & 767 \\
		$[20,30)$ & 24.1 & $[0.2,0.33)$ & 0.260 & 0.87 & 2531 \\
		$[30,40)$ & 34.4 & $[0.1,0.2)$ & 0.154 & 0.87 & 306 \\
		$[30,40)$ & 34.5 & $[0.2,0.33)$ & 0.259 & 0.87 & 940 \\
		$[40,55)$ & 46.3 & $[0.1,0.2)$ & 0.156 & 0.89 & 178 \\
		$[40,55)$ & 46.5 & $[0.2,0.33)$ & 0.259 & 0.88 & 449 \\
		$[55,140)$ & 73.2 & $[0.1,0.2)$ & 0.152 & 0.90 & 104 \\
		$[55,140)$ & 71.8 & $[0.2,0.33)$ & 0.257 & 0.88 & 295
	\end{tabular}
	\caption{Binning scheme for the \redmapper{} clusters and characteristics of the clusters in each bin, 
		including richness $\lambda$, redshift $z$ and probability of 
		correct centroid $\Pcen$.  The typical $N_{\mathrm{rand}}/N$ for the random catalogues
		is $\sim 22$, and the typical mean richness and redshift are consistent for the randoms
		to $\lesssim 1$ per cent.  All averages are weighted by the same lensing weights we use
		to generate stacked models.}\label{bintable}
\end{table*}

\subsection{Lensing data}
\label{sec:lenscat}

We use a shear catalogue first presented in \citet{2012MNRAS.425.2610R}, covering approximately 
9,000 square degrees and containing 39 million galaxies, or 1.2 galaxies/arcmin$^2$.  
This catalogue derives from Sloan Digital Sky Survey images as of Data Release 8
\citep{2000AJ....120.1579Y,2011ApJS..193...29A}.  The images were analyzed with the 
re-Gaussianization algorithm\footnote{An updated version of the software that was used to produce
  this catalogue is publicly available as
  part of the \texttt{GalSim} package \citep{2015A&C....10..121R}: \url{https://github.com/GalSim-developers/GalSim}.} \citep{2003MNRAS.343..459H}, which calculates adaptive second-order 
moments for the galaxy and point-spread function (PSF) by fitting elliptical Gaussians to the images, and then combines 
these moments and a correction for low-order non-Gaussianity to produce a measured distortion $e$.  We 
are interested in lensing shears, not distortions\footnote{The difference is merely a 
parameterization choice --- a shear is a ratio of linear functions of the axis ratio of the ellipse, 
while a distortion is a ratio of functions of the axis ratio squared.}, so we must correct for the 
difference, and also for the fact that shears do not add linearly \citep{2002AJ....123..583B}.  The 
average sensitivity of the mean distortion of the shape sample to an applied shear is usually called the responsivity,
which for unweighted measurements is $\mathcal{R} \approx 1-e_{\mathrm{rms}}^2$. For this catalogue the appropriate value of
$e_{\mathrm{rms}}$ is 0.365 \citep{2012MNRAS.425.2610R}.  We use the approximation here and not a full
calculation of $\mathcal{R}$, but our value is consistent with the more detailed analysis.  We also apply a
multiplicative shear calibration factor of 1.02, as discussed in \citet{2013MNRAS.432.1544M}.  Further 
characterization of the systematic errors in this catalogue was carried out in \citet{2012MNRAS.420.1518M} and
\citet{2013MNRAS.432.1544M}.

The photometric redshifts in this catalogue were calculated using the 
Zurich Extragalactic Bayesian Redshift Analyzer, or ZEBRA~\citep{2006MNRAS.372..565F}. 
ZEBRA is a 
template-fitting software; for this catalogue, four observed SEDs and two synthetic blue galaxy 
spectra from \citet{2000ApJ...536..571B} were used, along with twenty-five additional interpolated 
templates created between pairs of the six original templates.   The performance of the photometric
redshifts and their impact on weak lensing measurements was explored by \citet{2012MNRAS.420.3240N}.  The starburst-type galaxies, 
accounting for approximately 10 per cent of the original sample, were found to be unreliable based on 
comparison with a reference sample and were removed from the final catalogue.  There is a remaining 
known bias for galaxies with $z \ga 0.4$.  We know the true $\mathrm{d}N/\mathrm{d}z$ based on the work of 
\citealt{2012MNRAS.420.3240N}, so we can correct the effect of bias and scatter if we know the lens 
redshifts in the sample.  This bias is worse as redshift increases; it is 3 per cent for 
the distribution of cluster redshifts in our lower redshift bin ($z=0.1-0.2$) and 11 per cent for 
the distribution of cluster redshifts in our higher redshift bin ($z=0.2-0.33$).  All results shown
in this paper have these corrections applied.  The associated systematic uncertainty \red{on $\Delta\Sigma$} is 2
per cent.  \red{The uncertainty in the amplitude of the $\Delta\Sigma$ profile due to the use of 
photometric redshifts was estimated by comparing the different lensing amplitudes for the source 
population as estimated using representative spectroscopy from a variety of different spectroscopic
surveys.  The differences in the source redshift distribution between the different spectroscopic 
surveys reflects cosmic variance, and is the main limitation of our method.  For further details, 
we refer the reader to \citet{2012MNRAS.420.3240N}.}
 
\section{Mass modeling}\label{modeling}

In this section, we discuss the cluster mass model we will use to analyze the data.  
We exclude small scale data to minimize systematic uncertainties, particularly
with regards to membership dilution, strong shear, and galaxy obscuration by member galaxies.  The scales
we use are also large enough that the stellar mass component associated with the central galaxy is negligible, 
and can be safely ignored.   Consequently, we model the mass contribution from the halo profile only, without adding a component
for the central galaxy.
We first describe the model for a single cluster dark matter halo (section~\ref{subsec:nfw}) 
and then address the impact of analyzing multiple halos at once (section~\ref{subsec:stack}).  We summarize
the full set of parameters in section~\ref{subsec:likelihood} and describe how we will
constrain them using the data.

\subsection{The Lensing Profile of Cluster Halos}\label{subsec:nfw}

We assume that the clusters are spherical Navarro-Frenk-White \citep[NFW;][]{1997ApJ...490..493N} 
halos on average.  The mass of each cluster is assumed to depend on the cluster richness
via a scaling relation, including scatter.  The corresponding halo concentrations are computed
using a mass--concentration relation, whose amplitude we fit for.
In addition, some of the clusters are expected to be improperly centred.
We note that while no individual cluster is spherical, 
the observed lensing signal --- i.e. the tangential shear induced by
our galaxy clusters --- explicitly 
depends on the circularly symmetric mass density profile only.  That is, we only require
that NFWs be an adequate description of the circularly averaged projected mass density profiles.

Spherical NFW clusters have a mass density given by
\begin{equation}\label{NFW-rho}
\rho(r)=\frac{\rho_0}{(r/r_s)(1+r/r_s)^2}
\end{equation}
\citep{nfw96,1997ApJ...490..493N}, where $\rho_{0}$ and $r_s$ are two free parameters
that fully determine the density profile.  We measure cluster mass at a radius $r_{200m}$,
defined such that the mean density interior to this radius is 200 times the mean matter density
of the Universe at the redshift of the cluster.  That is,
\begin{equation}
\frac{M}{\frac{4\pi}{3}r_{200m}^3} = 200 \bar \rho_m(z).
\end{equation}
One also typically defines the concentration parameter $c = r_{200m}/r_s$.  The parameters $M$
and $c$ are uniquely related to $\rho_0$ and $r_s$, and are the parameters typically utilized to
characterize NFW halos.  We will follow that convention in this work.  

Our mass definition here is often referred to as $M_{200m}$ to denote the overdensity of 200 and the
fact that it is measured relative to the matter density.   For simplicity, we refer to $M$ or $M_0$
in some equations in this work, but this will always be $M_{200m}$.  Where a unit is needed
(such as in $\log M_0$), the masses have been measured in $\hMsun$ units.

As described in Section~\ref{bg-wl}, lensing shear is sensitive to a function of the surface mass 
density.  The surface mass density for NFW halos is analytic \citep{BartelmannNFW1996,WrightBrainerd},
so in the absence of centring errors we could plug this surface mass density directly into 
Eq.~\eqref{deltasigma} to fit our model to the data.  Centroiding errors 
introduce a further complication: they convolve the surface mass density with the 
distribution of centroiding offsets. Given an 
offset radius of $R'$, the convolution is~\citep{2006MNRAS.373.1159Y,2007ApJ...656...27J}
\begin{multline}
\Sigma(R, M, c, R') =  \\
  \frac{1}{2\pi}\int_0^{2\pi} \mathrm{d}\theta \,\Sigma(\sqrt{R^2+R'^2 + 2RR'\cos \theta}, M, c)
\end{multline}
We follow previous authors in adopting a 2D Gaussian miscentring distribution for $R'$
where we denote the Gaussian width as $R_{\textrm{mis}}$
\citep{2007ApJ...656...27J, 2016A&A...586A..43V}.  The fraction of well-centred 
clusters is denoted $\Pcen$.  The miscentred surface mass density is therefore
\begin{multline}\label{miscentring}
\Sigma(R, M, c, R_{\textrm{mis}}, \Pcen) = \\ 
  \Pcen\Sigma(R, M, c) \\
  +\frac{1-\Pcen}{2\pi}\int_0^{\infty}\mathrm{d}R' \left[\frac{R'}{R_{\textrm{mis}}^2} \exp(-R'^2/2R_{\textrm{mis}}^2) \right. \\
  \left.\times\int_0^{2\pi} \mathrm{d}\theta \,\Sigma(\sqrt{R^2+R'^2 + 2RR'\cos \theta}, M, c)\right]
\end{multline}

This expression is plugged into Eq.~\eqref{deltasigma} to obtain our final model for $\Delta\Sigma$:
$\Delta\Sigma(R, M, c, R_{\textrm{mis}}, \Pcen)$.
The natural length scale for parametrizing the cluster miscentring offset $R_{\rm{mis}}$ is the cluster radius $R_\lambda$
used to search for central galaxies.  Consequently, we set $R_{\rm{mis}} = \tau R_\lambda$, where $R_\lambda$
is the cluster radius from the \redmapper\ algorithm and $\tau$ is one of our model parameters.  If miscentring traces a satellite galaxy distribution 
modeled as a singular isothermal sphere,
one expects $\tau = \frac{1}{2}$.  For NFW halos, we expect $\tau < 1/2$.  Our exact priors
are detailed in Sec.~\ref{sec:miscentre}.

\subsection{The Mass--Richness Relation and Stacked Cluster Profiles}\label{subsec:stack}

In practice, we are not fitting to the lensing profile of individual clusters, but rather a stack of clusters of different 
redshifts and richnesses.  The model we fit to the data is built in a similar way, by stacking the 
expected lensing signal given the richnesses and redshifts of the clusters that went into the
measured lensing signal, along with their miscentring information.
The relation between halo mass and cluster richness is 
given by the probability distribution $P(M|\lambda)$,
for which we adopt a log-normal model.  We set the mean of this relation via
\begin{equation}\label{massmodel}
 \avg{M|\lambda} = M_0 \left(\frac{\lambda}{\lambda_0}\right)^{\alpha} 
\end{equation}
using a pivot point of $\lambda_0 = 40$.  This choice roughly de-correlates our model parameters of 
scientific interest.
In our fiducial model we do not allow for redshift evolution.  An extension of our fiducial model that allows for
redshift evolution in the scaling relation is presented in Appendix~\ref{app:redshift}, where we demonstrate
that the data has only minimal constraining power on redshift evolution, and 
that allowing this additional degree of freedom does not impact the conclusions of this work.

We note that the above parameterization of the richness--mass relation differs from the more traditional
convention of defining the scaling relation parameters via $\avg{\ln M|\lambda} = \ln M_0 + \alpha \ln (\lambda/\lambda_0)$.
The reason is that unlike the traditional parameterization, Eq.~\ref{massmodel} effectively decouples
uncertainty in the scatter from uncertainty in the amplitude of the mass--richness relation.  
If desired, one can go from one choice of parameterization to the other via
$\ln \avg{M|\lambda} = \avg{\ln M|\lambda} + 0.5\Var(M|\lambda)$ \citep{rozoetal09a}. 

The variance of the distribution $P(M|\lambda)$ is modeled based on our expectation for the scatter
in the converse distribution $P(\lambda|M)$.  Specifically, consider a model in which the scatter in $\lambda$
is the sum in quadrature of a Poisson term and an additional intrinsic variance term,
\begin{equation}
\Var(\lambda|M) = \frac{1}{\avg{\lambda|M}} + \sigma^2_{\ln \lambda|M}.
\end{equation}
Using the framework of 
\citet{2014MNRAS.441.3562E}\footnote{To address a possible source of confusion, the
  \citet{2014MNRAS.441.3562E} framework uses $\alpha$ to denote the power-law slope of parameters
  given mass; i.e. to zeroth order, our $\alpha$, denoting the power-law slope of mass given $\lambda$, 
  is the inverse of theirs.}, we can readily invert this expression to arrive at
\begin{equation}
\Var(\ln M|\lambda) = \frac{\alpha^2}{\avg{\lambda|M}}  + \alpha^2\sigma^2_{\ln \lambda|M}.
\end{equation}
This suggests that we model the variance in $\ln M$ at fixed richness via
\begin{equation}
\Var(\ln M|\lambda) = \frac{\alpha^2}{\lambda} + \sigma^2_{\ln M|\lambda}
\end{equation}
\red{where we have changed the parametrization to match the type of scatter we have measured.}

\citet{rozorykoff14} and \citet{rozoetal15} have estimated the scatter in mass at fixed richness by comparing the \redmapper\ catalogue
to existing X-ray catalogues and to the {\it Planck} SZ cluster catalogue \citep{2014A&A...571A..29P}.  
We  summarize their findings as $\sigma_{\ln M|\lambda} = 0.25 \pm 0.05$. 
This means that the Poisson term dominates the scatter for our lowest richness 
bin, while the additional intrinsic scatter in the $M-\lambda$ relation dominates in our upper richness bins.  
We tested both Gaussian and lognormal scatter models and find no difference in 
the recovered
parameter values for the mass--richness relation.
We adopt the lognormal model as our fiducial model as that is the standard in the literature.

We must also weight the cluster models appropriately in the stack.  Clusters at different redshifts 
have different lensing efficiencies for background galaxies at different redshifts, and the background galaxies themselves 
contribute less information to the lensing signal as they become fainter and harder to measure; also, we measure the lensing signal 
using physical distances in the plane of the lens, 
so clusters with a larger angular 
diameter distance will contribute fewer pairs to the lensing signal for a fixed source number density in 
angular coordinates.  We construct a per-lens weight that is a function of the angular diameter
distance to the lens (which accounts for the physical aperture) as well as the average weight applied to
background galaxies (Equation~\ref{Weights})
for each lens, which accounts for both the source galaxy redshift distribution and for the fact that the average
source galaxy shape weight is a function of redshift (due to the size of the measurement error).

We have corrected our lensing signal for the fact that we have imperfect, biased knowledge of the source
\photoz{}s.  This means we should use the true redshift distribution $P(z_s)$, which we take to be the
parametric curve of \citet{2012MNRAS.420.3240N}, rather than our biased measured distribution,
to compute the expected average weights.  If we also had perfect knowledge of the
source shape weights as a function of true redshift (and not measured \photoz), we would simply compute
\begin{equation}
\langle w_s \rangle(z_l) = \int_0^{\infty} \mathrm{d}z_s\ P(z_s) \Sigma^{-2}_{\rm cr}(z_l, z_s)w_{\rm shape}(z_s).
\end{equation}
While we do not have such knowledge, the weight is only a slowly varying function of $z_s$, with $\sim 5$~per cent
change over the redshift range covered by our source catalogue, so even with our biased \photoz{}s, a direct
measurement of $w_{\rm shape}(z_s)$ from our catalogue should contribute negligible errors to our final model prediction.
We measure the average shape weights in redshift bins $j$ (which we choose to be narrow bins of width 0.01 in redshift) 
in our catalogue and combine them with the true redshift 
distribution to obtain
\begin{equation}
\langle w_s \rangle(z_l) =\sum_j \Delta z_S P_\mathrm{true}(z_s) \Sigma^{-2}_{\rm cr}(z_l, z_s) \langle w_{\mathrm{shape}}(z_{s,j})\rangle.
\end{equation}
Finally, once we account for the differing aperture sizes on the sky for lenses at different redshifts, we obtain a per-lens weight of
\begin{equation}
w_l(z_l) = \frac{\langle w_s \rangle(z_l)}{D_A^2(z_l)},
\end{equation}
normalized so the sum of weights over all clusters is 1.

Putting it all together,
given clusters $m$ divided into bins $n$ with redshifts $z_m$ and richnesses $\lambda_m$, 
the expectation value for the lensing profile of a cluster stack is
\begin{equation}\label{deltasigmamodel}
\avg{\Delta\Sigma(R)}_n = \sum_{m\in n} w(z_m) \Delta\Sigma(R|M_m,c_m,\Pcen,R_{\rm{mis}}).
\end{equation}
When building the model stack, the mass $M_m$ of cluster $m$ is drawn from the probability
distribution $P(M|\lambda_m)$ including the aforementioned scatter model.
The concentration $c_m$ is then drawn from a distribution with lognormal width 0.14 dex \citep{2006ApJ...652...71W} and 
a mean set according to the mass--concentration relation 
of \citet{2013ApJ...766...32B} computed by the \texttt{colossus} package \citep{2015ApJ...799..108D}.
We allow
for an overall amplitude shift of this fiducial relation 
by a multiplicative constant $c_0$.
We find that including scatter in the mass--concentration relation has a  negligible impact on our results, but we include it
in accordance to theoretical expectations.
For further details and alternative mass--concentration relations, see Appendix~\ref{app:models}.

\subsection{Likelihood Model}\label{subsec:likelihood}

We compute the stacked weak lensing signal of \redmapper\ clusters in 4 bins of richness $\lambda$
and 2 redshift bins.  The characteristics of the bins are shown in Table~\ref{bintable}.  
For each galaxy cluster, we compute the observed lensing profile $\widehat{\Delta\Sigma}$ as per Eq.~\eqref{DeltaSigmaEstimatorStack}, as well
as the theoretical model $\Delta\Sigma$ as per Eq.~\eqref{deltasigmamodel}.
In computing our theoretical prediction, the $M$ assigned to each cluster is scattered relative to its expectation
value to take into account the scatter in the mass--richness relation.  
Given our Monte Carlo approach, this is formally equivalent to treating the mass of each individual cluster as a free parameter,
and marginalizing over all these parameters as part of our MCMC.
That is, our procedure is equivalent to a Monte Carlo evaluation of the appropriate
integral.
We truncate the richness measurements at $\lambda=140$ to avoid introducing a sparsely populated 
bin of extremely rich galaxy clusters, where the predicted profiles are unstable due to the random
realization of the scatter.

With the observed and modeled lensing profiles, we model the likelihood function as a Gaussian
$\lkhd \propto \exp( -\frac{1}{2}\chi^2 )$ where
\begin{equation}
\chi^2 = \left[ \widehat{ \Delta\Sigma}(R) - \avg{\Delta\Sigma(R)} \right] \mathbf{C}^{-1}  \left[ \widehat{ \Delta\Sigma}(R) - \avg{\Delta\Sigma(R)} \right].
\end{equation}
The covariance matrix $\mathbf{C}$ is, for this analysis, diagonal, with the diagonal terms
computed as the variance on the mean $\Delta\Sigma$ for each bin.  As weak lensing is dominated on
small scales by
the intrinsic shape distribution of galaxies (``shape noise''), with each source contributing for
one lens on average, covariances between bins are typically very small.  Jackknife error bars are usually used to estimate this effect, but for these scales,
the noise in the jackknife is significantly larger than the expected value of the off-diagonal terms
and has a detectable impact even on the diagonal terms, so we use the variance on the mean
directly. Based on previous work with this SDSS shape catalogue, we expect this
approximation to work well for $R<5$ Mpc, outside our upper limits for our fitting range 
\citep{2013MNRAS.432.1544M}.

We choose our radial range to avoid contamination from nearby large-scale structure (the so-called 
``two-halo term'' of halo modeling) and to avoid problems due to background selection and increased
scatter due to low sky area in the inner regions of clusters.  We use $0.3\ h^{-1}$ Mpc as the interior
radius limit.  \citet{2010MNRAS.405.2078M} suggest using minimum scales of 15-25 per cent of the virial radius;
our lensing-weighted average virial radius, given the parameters we find below, is about $1.3\ h^{-1}$~Mpc,
so our lower radius limit corresponds to $\sim R_{\rm vir}/4$.  
We choose a richness-dependent upper limit of $2.5 \left(\lambda/20\right)^{1/3} h^{-1}$~Mpc based on a comparison of 
halo model predictions to the single-halo measurements from the simulations we will use for validation
in Sec.~\ref{sec:sims}.  We also tested values of the constant in front of the $(\lambda/20)^{1/3}$
factor above ranging from $1$ to $7$, and found no 
statistically significant change in the scientific parameters of interest.

Finally, as will be discussed in Sec.~\ref{sec:sys}, we have an overall calibration 
uncertainty in our $\Delta\Sigma$ values due to possible shear estimation biases and photo-$z$ biases. 
To marginalize over these effects, we include in our theoretical model a parameter $b$ such that
\begin{equation}
\widehat{\Delta\Sigma_b}(R) = (1+b)\widehat{\Delta\Sigma}(R).
\end{equation}
We assume that all clusters are equally affected by this systematic, which maximizes the impact of
these effects on the amplitude of the mass--richness relation (i.e. all clusters move up and down in
unison).

Our final set of seven model parameters includes the scaling relation parameters
$\log_{10} M_0$, $\alpha$, $\sigma_{\ln M|\lambda}$, 
the miscentring parameters $\Pcen$ and $\tau$, 
the multiplicative constant $c_0$ that rescales the mass--concentration relation
of halos relative to our fiducial model, and the overall multiplicative amplitude shift
$1+b$ corresponding to possible systematic errors in the lensing signal measurement.
\red{We use informative priors on parameters that are difficult to measure from the data
due to degeneracies ($\sigma_{\ln M|\lambda}$, $\Pcen$, $\tau$, and $b$) and implicitly
fix one parameter with strong disagreement between the data and the prior due to our belief
that the data is incorrect (the redshift evolution of the mass-richness relation).} 
We use the \texttt{emcee} package\footnote{\url{http://dan.iel.fm/emcee/}} \citep{2013PASP..125..306F} 
to sample our likelihood function with 100
walkers.   We discard the first 100 steps of each walker as burn-in steps.  Our reported values correspond to the 
median result of all samples, and our errors 
correspond to the difference between the median and the 16th or 84th percentiles of the samples.

\section{Systematic Uncertainties}\label{sec:sys}

\subsection{Measurement Systematics}

\red{In this section, we consider systematic uncertainties in the measurement process, including
  shear and photometric redshift systematics, and systematics associated with our estimator for the
  lensing signal.}

The shear catalogue we use was extensively tested in \citet{2012MNRAS.425.2610R}, with further
error characterization in \citet{2013MNRAS.432.1544M}. Based on that work, taken together, we expect 
a systematic error budget of 5 per cent on $\Delta\Sigma$ for lenses
in this redshift range and our galaxy shape sample, comprising errors from PSF correction,
noise bias, selection bias, and photo-$z$ biases.  The first three items in that list, which are linked and 
which all cause errors
or biases in shear estimation, were measured in realistic galaxy simulations and were found to contribute 3.5 per 
cent to this error budget.  The remnant of the systematic error is dominated by \photoz{} biases, as measured by
\citet{2012MNRAS.420.3240N} through comparison to spectroscopic samples.  Other effects were also considered,
including, for example, stellar contamination and shear responsivity errors, but all were found to be
subpercent and thus subdominant to the errors mentioned above.
Additionally, the calibration of the recovered shears was tested up to an
induced shear of 0.1, larger than we expect to find even for the clusters in our most massive bin at a radius of 
$0.3 h^{-1} $ Mpc, so we should not see biases from mismeasured larger shears near the cluster centres.  We
refer interested readers to \citet{2012MNRAS.425.2610R}, \citet{2012MNRAS.420.1518M}, and 
\citet{2013MNRAS.432.1544M} for more information.

In addition to shear and photo-$z$ biases, our recovered cluster masses and concentrations can 
be affected by magnification and
obscuration\footnote{Here, obscuration refers not only to
the literal blocking of source galaxies by member galaxies, but also the inability to properly
select them due to blending with member galaxies.  \red{This is sometimes called crowding.}} of the background sky by cluster member galaxies.  However, 
for this SDSS shape sample, the low-redshift \redmapper\ clusters we are using, 
and the range of transverse separations considered here,
these effects are negligible, as demonstrated in 
\citet{2015MNRAS.449.1259S}.  
The slope of the background galaxy population number counts with flux and
size, which contribute to the size of the magnification effect \citep{2009PhRvL.103e1301S}, 
are too shallow to make magnification detectable for this lens sample and range of $R$.
Obscuration effects are
large when many of the cluster galaxies can be seen (deeper imaging), leading to portions of the background sky
being unobservable either directly or through blending. Again, this is less pronounced in
SDSS than in deeper surveys.  We measure number densities to
determine the obscuration area of the \redmapper\ member galaxies, and rerun the fitting code
from \citealt{2015MNRAS.449.1259S} for our much larger radius range. We find 
sub-percent biases in our final mass determinations.  Consequently, we ignore magnification
and obscuration as sources of systematic uncertainty in our measurements.
    
As a further test for systematic errors due to observational effects such as shear and photometric
redshift estimation, we compare our measurements to those from a fully independent shear
and photo-$z$ catalogue.  The
catalogue, which we will refer to as the \scatname\ catalogue, has been used in a
number of other lensing studies
\citep[e.g.,][]{MelchiorVoids2014,2014ClampittVoids}.  The shapes in
the \scatname\ catalogue were also measured using the \regauss\ technique, but
the code was developed independently. The code used to generate the catalogue is
freely available online\footnote{https://github.com/esheldon/admom}.  The
\scatname\ catalogue also makes use of the \photoz s released by
\citet{SheldonPhotoz2012}.  The full photometric redshift probability
distributions \pofz\  were used to calculate the \dsig\ for each lens-source
pair.  \dsig\ was measured using the publicly available \xshear.\footnote{https://github.com/esheldon/xshear}

Because the same \regauss\ PSF-correction algorithm was used for both our primary catalogue
and the \scatname\ catalogue, we may expect the catalogues to have the same overall
shear calibration.  However, it is still useful to compare these catalogues.
First, there could be real differences in the implementation of the \regauss\
algorithm, resulting in small differences in the final shear calibration.
Second, differences in the \photoz\ generation and use of the full \pofz, rather
than the point estimates with correction for resultant biases, could result in different overall normalization
for \dsig\ if done inconsistently.  Third, although the same data were used as input
for both codes, how the data were used and organized, and the galaxy selection criteria, differs in detail.  
Fourth, there could be software ``bugs'' that could result in inconsistent lensing results 
between the catalogues.  

\begin{figure}
	\includegraphics[width=0.45\textwidth]{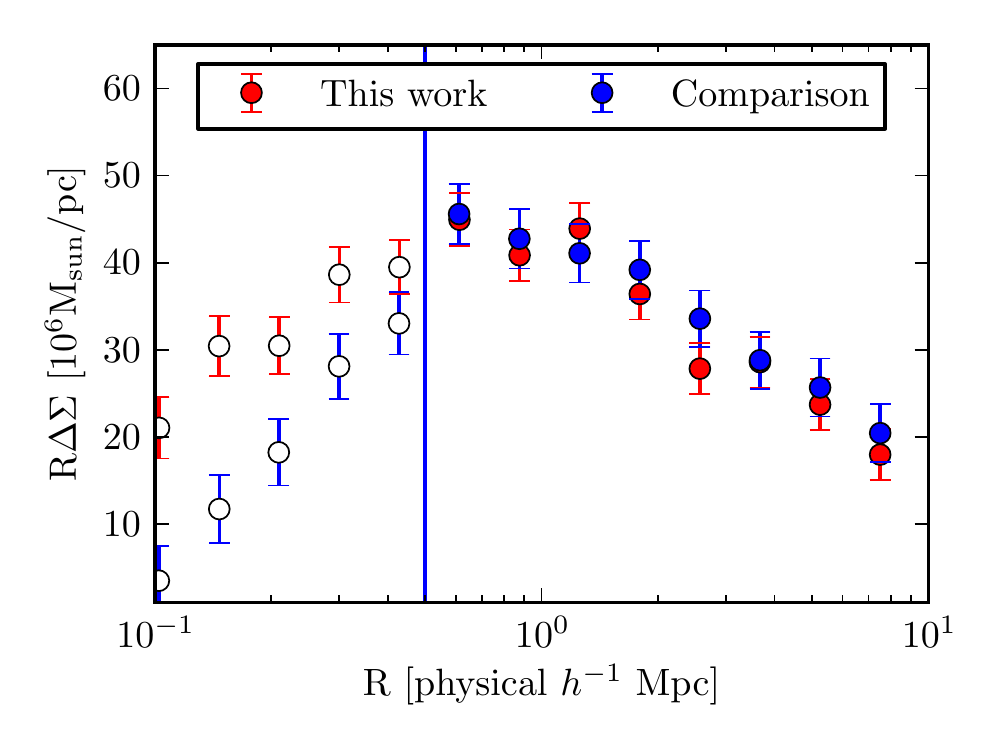}
	\caption{Lensing signal from the \redmapper{} catalogue for our analysis (``this work'') and the
      comparison \scatname\ catalogue (``comparison'').  We consider the signals comparable at all
      scales $\ge 500h^{-1}$~kpc.  Filled circles are comparable data points, while open circles
      are shown for scales smaller than the cutoff.
        }\label{shearcomp}
\end{figure}

We show the signal for \redmapper{} clusters with redshifts $0.1-0.3$ and richnesses $\lambda=20-203$
in Figure~\ref{shearcomp}.  We compare the radial scales $500 h^{-1}$~kpc to $10 h^{-1}$~Mpc. The lower edge is
set by the presence of selection effects in the \scatname\ catalogue 
that lead to scale-dependent suppression of the
lensing signal.  We note that these selection effects were known {\it a priori}, but that no attempt was made to
correct for these since the goal of this test is simply to compare the \dsig\ calibrations of the two independent shear and
photo-$z$ pipelines in the regime where selection effects are unimportant.  The upper limit of the scales used in our tests is set by the maximum
radius considered in this work.
We test for consistency between the two shape and photometric redshift catalogues by minimizing
$\chi^2 = \sum[\Delta\Sigma' - (1+a)\Delta\Sigma]^2/\sigma_{\Delta\Sigma}^2$
for some constant $a$. We find $a=0.031\pm 0.033$, largely insensitive to the magnitude of the
correlations between the two $\Delta\Sigma$ estimators, and also insensitive to the exact
end points used for the comparison as long as we are above the 500$h^{-1}$~kpc cutoff.
Thus we can exclude any significant relative differences in calibration of $\Delta\Sigma$ between these two 
independently-produced catalogues, in line with our expectations of $\lesssim 5$ per cent multiplicative
bias as detailed above.

In our work, we have corrected for dilution of the source sample by cluster member galaxies under the assumption that cluster member galaxies
have zero shear.  If cluster member galaxies are radially aligned, however, \red{ our correction will be underestimated}.  \citet{2012JCAP...05..041B} tried to measure the radial intrinsic alignment (IA)
signal of galaxies in our source sample with respect to the positions of Luminous Red Galaxies (LRGs) at a similar redshift to our clusters.  They were
unable to detect an intrinsic alignment signal, placing an upper limit of 1.5\% \red{on the
  fractional contamination of intrinsic alignments}
on the recovered galaxy-galaxy lensing signal \red{at a separation of $1h^{-1}$Mpc}.  \red{To
  estimate an upper limit for this work, we must consider two factors: differences in the intrinsic
  alignment strength, and differences in the $\Delta\Sigma$ (which will modify the fractional
  contamination estimate).  To estimate the former, we use the fact that the intrinsic alignment
strength depends on the type and luminosity of the source galaxies (which is the same in this work
and that one), and on the density tracers used as lenses.  \citet{2015MNRAS.450.2195S} provides an
empirically-determined power-law relationship between the small-scale intrinsic alignment strength
and the halo mass of the density tracer, which we can use to scale up the estimated intrinsic
alignment.  After taking} 
into account the increase in lensing signal $\Delta\Sigma$ between our clusters and LRGs, we find
that the 1.5\% upper limit on fractional contamination for LRG lenses corresponds to an $\approx 6\%$ upper limit for galaxy clusters.
While this upper limit is comparable to our uncertainties, we note that IA remains undetected for
typical \red{satellite} galaxy populations in 
galaxy clusters \citep[e.g.,][]{2015A&A...575A..48S}, with detections only for 
LRGs \citep[e.g.,][]{2015MNRAS.450.2195S}.   For the purposes of this work, we have chosen 
not to increase our systematic error budget, pending future investigations of intrinsic alignments 
in galaxy clusters.  Indeed, these arguments suggest that IA may soon become an important systematics
that must be simultaneously fit for when analyzing stacked cluster lensing data \red{(unless all galaxies at
or very near the cluster redshift can be robustly removed)}.

\subsection{Cluster Miscentring}
\label{sec:miscentre}

Some fraction of the \redmapper\ clusters are expected to be miscentred.  Based on the \redmapper\ miscentring probabilities,
we expect 87\% of the galaxy clusters to be correctly centred.  This value is consistent with the fraction of clusters for which
the X-ray centre of the galaxy cluster is in close proximity to the central galaxy assigned by
\redmapper\ \citep{rozorykoff14}.
Here, we further characterize the miscentring scale $\tau$ that governs the radial offset of miscentred clusters.
Specifically, we compute the distribution of positional offsets $R$ of galaxy clusters shared between \redmapper\
and two high-resolution X-ray external data sets, the XCS DR1 \citep{2012MNRAS.423.1024M} and ACCEPT \citep{2009ApJS..182...12C} cluster catalogues.
The distributions are modeled as
\begin{equation}
P(R) = \frac{f}{R_0}\exp(-R/R_0) + (1-f)\frac{R}{\tau^2 R_\lambda^2}\exp\left( -\frac{1}{2}\frac{R^2}{\tau^2R_\lambda^2} \right).
\end{equation}
This distribution has 3 free parameters: the fraction of clusters that are correctly centred $f$, 
the parameter $\tau$ governing the centring offset distribution of miscentred clusters,
and the parameter $R_0$ describing
the centring offset between the external data set and \redmapper\ clusters for correctly centred
clusters.  Note that correctly centred clusters will still have offsets relative to external data sets due to both
centring uncertainties in the external data set and physical displacements
between the location of the central galaxy and the centre of the intra-cluster gas.  \red{The 
parameter $R_0$ accounts for these displacements.  It is usually small---such offsets tend to peak 
at $\lesssim 0.1$ Mpc \citep{2016MNRAS.456.2566C}.  As $R_0$ models an offset in the X-ray 
detection relative to the dark matter halo, which is not relevant for our optical centring 
measurements here, we do not include it in our model width. Additionally, since the parameter 
itself is smaller than our uncertainty on our prior, any errors in $R_0$ should already be reflected in our modeling.}
  
  The likelihood of each of the combined data sets (XCS and ACCEPT) is the product
of the probability $P(R)$ over the clusters in the joint cluster samples.  The resulting constraints are summarized 
in Table~\ref{tab:miscentre}, where we report the mean and standard deviation of the parameters computed from
the MCMC.  We do not report on the posterior of the parameter $R_0$ since it is purely a nuisance
parameter for this study. 

\begin{table}
\centering
\begin{tabular}{c l c c }
External Data Set & Number & $f$ & $\tau$\\
\hline
XCS & 82 & $0.85 \pm 0.05$ & $0.48 \pm 0.09$ \\
ACCEPT & 54 & $0.75 \pm 0.06$ & $0.31 \pm 0.05$ \\
\end{tabular}
\caption{Constraints on the miscentring parameters from each of the
various external data sets.  The ``number'' column refers to the total
number of clusters in the sample. The parameter $f$ is the fraction of
miscentred clusters; the parameter $\tau$ characterizes the width of
the centring offset distribution of miscentred clusters.}\label{tab:miscentre}
\end{table}

From Table~\ref{tab:miscentre}, we see that the two high-resolution
X-ray data sets are consistent with each other and with our
expectation $f=87\%$.  We consider two analyses: one in which the
miscentring priors are chosen following a ``middle-of-the-road''
approach, with a gaussian prior $f=0.80 \pm 0.07$ and $\tau = 0.40 \pm
0.1$, and one for which the miscentring fraction is set to the
expected value $f=0.87$ based on the \redmapper\ centring
probabilities.  We note that from the summary of our binning scheme in
Table~\ref{bintable} one can see that the reported fraction $f$ from
the \redmapper\ algorithm itself is roughly constant for all $\lambda$
and redshift bins, so we do not include further richness or redshift
evolution in this parameterization beyond what is implied by the
dependence of miscentring radius on the cluster radius.

\subsection{Cluster Projections}
\label{sec:proj}

Photometric cluster finding is known to suffer from projection effects.  That is, two neighboring clusters that fall along the line
of sight will be blended by the photometric cluster finder into a single, apparently larger cluster.   We estimate the projection
rate in the \redmapper\ cluster catalogue to be $p=10 \pm 4$ per cent (see Appendix~\ref{app:proj}).  Characterizing the impact of such projections on cluster mass calibration
is not trivial, but rough scalings can be readily estimated.  
In particular, clusters that appear in projection will also have a lensing signal affected by said projection effects.
If the richness $\lambda$ scales roughly linearly with the mass $M$, then the total projected mass per unit richness will remain 
constant, implying that projected clusters simply slide up and down the richness mass relation, without deviating from it.
Indeed, this effect has been seen in numerical simulations \citep{anguloetal12,nohcohn12}.
Below, we quantify this effect in a way that can be incorporated into our analysis.

Consider a projection of two halos of mass $M$ and richness $\lambda$ which are blended into a single cluster of richness
$2\lambda$.  Assuming $M\propto \lambda$, a clean cluster of richness $2\lambda$ would have a mass $2M$.
By contrast, the blended cluster will appear to have a mass $(1+\epsilon)M$, where $\epsilon$ characterizes the effective 
mass contribution of the projected halo.  This effective mass will depend on the relative position of the two blended halos along
the line-of-sight, the concentrations of the two halos, etc., but we will treat $\epsilon$ as a single effective parameter
that characterizes all of these effects.   Letting $\avg{M}_0$ be the mass of galaxy clusters of a given richness in the absence
of projection effects, and $p$ be the fraction of projected clusters,
then the observed weak lensing mass of a cluster stack will take the form
\begin{equation}
\avg{M} = (1-p)\avg{M}_0 + p(0.5+\epsilon) \avg{M}_0
\end{equation}
Note that projected halos will only contribute their full mass ($\epsilon=1/2$) if they are perfectly aligned,
and will never contribute no mass ($\epsilon=0$).  We generously set $\epsilon = 0.25 \pm 0.15$,
which encompasses the two previously mentioned extremes at less than $2\sigma$. 
Solving for $\avg{M}_0$, we finally arrive at
\begin{equation}
\frac{\avg{M}_0}{\avg{M}} = \frac{1}{1+p(\epsilon-0.5)}  = 1.02 \pm 0.02.
\end{equation}
The uncertainty $\pm 0.02$ was estimated by randomly drawing both the fraction of projected clusters $p=0.10 \pm 0.04$,
and the parameter $\epsilon=0.25 \pm 0.15$.  We computed the ratio $\avg{M}_0/\avg{M}$ for each realization, and repeated
$10^4$ times to estimate the corresponding variance in the correction factor.

\subsection{Cluster Triaxiality}\label{sec:triaxial}

Dark matter halos are known to be triaxial.  Optically selected halos are expected to be biased
so that they are preferentially elongated along the line of sight.  In this case, the galaxy contrast
relative to the immediate neighborhood should be maximized, making cluster detection easier.
The elongation of the halo along the line of sight will naturally lead to an enhanced weak lensing
signal, and so the recovered mass--richness relation may well be affected by this type of selection
effect.  For our purposes, the key point is that cluster triaxiality induces covariance between
cluster richness and weak lensing mass estimates. 

The magnitude of this type of selection effect can be readily estimated using the multi-model
cluster component of \citet{2014MNRAS.441.3562E}.  If $\avg{M}_0$ is the halo mass 
in the absence of triaxiality-induced selection effects, and $\avg{M}$ is the recovered weak lensing 
mass, then one has
\begin{equation}
\frac{\avg{M}_0}{\avg{M}} = \exp\left[ - \beta r \sigma_{\ln M|\lambda}\sigma_{\ln M|WL} \right]
\end{equation}
where $\beta$ is the slope of the halo mass function at the relevant scale, $r$ is the correlation
coefficient, and $\sigma_{\ln M|\lambda}$ and $\sigma_{\ln M|WL}$ is the scatter in mass at fixed
richness and weak lensing mass respectively.  Using the simulation results of \citet{nohcohn12},
we find $r=0.25$.  We do not have an uncertainty estimate for this correlation coefficient,
so we adopt a prior $r\in[0,0.5]$. 
We set $\sigma_{\ln M|WL}=0.25 \pm 0.05$ based on the results by \citet{beckerkravtsov11},
and $\sigma_{\ln M|\lambda}=0.25\pm 0.05$ as discussed earlier.  We use a Monte Carlo approach
in which all of the above quantities are randomly drawn in order to estimate the final correction
and its uncertainty. We arrive at
\begin{equation}
\frac{\avg{M}_0}{\avg{M}} = 0.96 \pm 0.02
\end{equation}
That is, weak lensing masses overestimate the true mass of photometrically selected clusters by $4\%\pm 2\%$.
This value is in excellent agreement with the simulation-based estimates of \citet{dietrichetal14}.

The combined effects of cluster projections and cluster triaxiality can be summarized as
\begin{equation}
\frac{\avg{M}_0}{\avg{M}} = 0.98 \pm 0.03
\end{equation}
We apply this multiplicative correction to our recovered best-fit amplitude of the mass--richness relation {\it a posteriori},
and add in quadrature
the above 3\% systematic uncertainty to our total systematic error budget.

\subsection{Modeling Systematics}\label{sec:sims}

Recent work indicates that the more flexible Einasto profile \citep[3
free parameters;][]{Einasto,duttonmaccio14} may describe dark matter
halos more accurately than our assumed NFW profile.  Depending on the
details of the weak lensing analysis, the differences between Einasto
and NFW profiles can be significant.  For instance,
\citet{2016JCAP...01..042S} find biases from fitting NFW profiles to
Einasto halos that range from -1\% for low- and middling-mass clusters
to $\sim +15\%$ for the highest-mass clusters. However, these biases
come from fits that use smaller radii (by a roughly a factor of
two) than the smallest radial bin considered in this study.  The
biases we see should be less, since the difference between NFW and
Einasto profiles is largest at small scales;
\citet{2008JCAP...08..006M} performed fits over a radius range more
similar to ours and found negligible differences in the masses and
moderate differences in the concentrations. Therefore, we do not
expect to see such large biases because of the modest range of scales
we probe.  Nevertheless, the main point stands: one should check
whether the profile assumed induces a significant bias in the weak
lensing mass estimate.  This is especially true in our case, since
we rely on an NFW halo out to scales comparable to the splashback
radii of our halos, where one expects systematic deviations from the
NFW profile \citep{2014ApJ...789....1D}.

Here, we address this source of systematic uncertainty using numerical
simulations to test whether our parametric models can introduce
significant biases in our recovered weak lensing masses.
Specifically, we use an $N$-body simulation with a volume $V=[1.05\
h^{-1}\,{\rm Gpc}]^3$ and 2.7 billion particles, run with the
\textsc{L-Gadget} code, a variant of \textsc{Gadget}
\citep{Springel2005}.  The cosmological model is flat $\Lambda$CDM
with matter density $\Omega_m=0.318$, $\sigma_8= 0.835$, and $h=0.670$,
and $n_s=0.961$. 
We use the halo catalogue generated by the \textsc{Rockstar} halo
finder \citep{Behroozi2013a}.  In this simulation a $3.5 \times
10^{13}\ h^{-1} \msun$ halo is resolved with $10^3$ particles.

We construct synthetic weak lensing profiles from the numerical simulation.  First, we
divide the simulation into 64 jackknife regions.  We
assign an observed ``richness'' to each halo in the simulation via 
$\lambda=M_{\rm{obs}}/10^{14} h^{-1} M_{\odot}$.  $M_{\rm{obs}}$ is the effective observed
mass of the halo, computed via $\ln M_{\rm{obs}} = \ln M_{\rm{true}} +\delta$ where $\delta$
is a Gaussian random draw of zero mean and variance $\avg{\delta^2}^{1/2}=0.2$, intended to
model lognormal scatter in the mass--observable relation.
Clusters are sorted into richness bins, and the lensing profiles for these bins are fit using our 
likelihood model with no miscentring.  We change the $\Omega_m$ of our model-fitting pipeline 
to the value from the simulations for these tests only.
We adopt the same mass--richness relation model as Eq.~\eqref{massmodel}, this time with pivot
$\lambda=1$, and we allow for lognormal scatter in mass at fixed richness. We use the data
error bars to replicate our sensitivity to different radius scales, and combine the results from
the 64 jackknife samples to measure any potential bias at a higher precision than the data 
error bars would allow from any one fitting procedure alone.

 Using the formalism in \citet{2014MNRAS.441.3562E},
the resulting best fit $M_0$ value should be 
$\log M_0 = 14.0 - \beta\sigma_{M|\lambda}^2/\ln(10) = 13.953 \pm 0.006$, where $\beta$
is the slope of the halo mass function over the range of halo masses probed.
The error bar in our theoretical predictions comes from the difference between the first and second-order
calculation using the formalism of \citet{2014MNRAS.441.3562E}.
We compare this expectation against our recovered $M_0$ values to characterize
the systematic uncertainty $\Delta \log M_0$ associated with our parametric modeling.
We find results consistent with our expected value, \red{$13.953 \pm 0.001$ if we use all 
simulation mass bins or $13.943 \pm 0.004$ if we use only the bins within our expected
observational mass range}.  As the error bar is dominated 
by the theoretical uncertainty, we choose to add no further
uncertainty to our error budget based on this comparison.

\subsection{Baryonic effects}

Baryonic physics may modify the matter density profile of halos relative to that observed in dark matter-only
simulations.  Simulations that include baryonic cooling as well as AGN feedback have been used to study the
impact of baryons on halo profiles~\citep[e.g.,][]{schaller15,2016arXiv160206668C,bocquet16}.  We summarize
the trends in these papers as follows.  First, on very small scales, the stellar mass component of the central galaxy
dominates.  These scales are excluded from our analysis.  At somewhat larger scales, the profiles become either
more or less concentrated depending on the relative impact of baryonic cooling to AGN feedback.  The halos
are still roughly NFW, but the scale radius $r_s$ changes, reflecting the overall mass redistribution within a halo.
The fact that we marginalize over a rescaling of the concentration--mass relation should allow us to
avoid substantial biases in mass estimates due to the change in halo concentrations.  
Third, and most importantly for our purposes, the mass within $R_{200m}$ is very stable, well below the $5\%$
level.  As such, we expect the impact of baryons is well below our final error budget, and is therefore ignorable.
Nevertheless, we caution that the impact that baryonic physics has on $M_{200m}$ is likely to become relevant to 
future weak lensing experiments.  Fortunately, a straightforward generalization of
the calibration program detailed above relying on simulations with baryonic physics should easily
allow us to incorporate such systematics in future analyses.

\subsection{Systematics Summary}

The systematics we consider are: shear and source photometric redshift systematics ($\pm 5\%$ in $\Delta\Sigma$),
cluster projections, halo triaxiality (combined $-2 \pm 3\%$ in $M_0$), and modeling systematics (no additional uncertainty).
We handle the $\Delta\Sigma$  errors via marginalizing over the fitting parameter $b$ as described in
Section~\ref{subsec:likelihood}; we note that this is a 5 per cent tophat prior, which when combined with other
Gaussian errors should contribute approximately 3.5 per cent to a Gaussian uncertainty on the overall amplitude.
We apply the errors and bias correction in $M_0$ due to halo triaxiality and halo projections
{\it a posteriori} to our measured mass amplitude from our fitting procedure.   

In the previous sections, we specified miscentring priors, but we did not estimate how miscentring impacted
the uncertainty in the recovered weak lensing mass.  To test this, we compare the results of our fiducial analysis
to a second analysis in which the miscentring parameters are held fixed at their fiducial values.
We find the resulting uncertainties in the amplitude of the mass--richness relation are essentially identical, so
cluster miscentring does not appear to have a significant impact on the precision of weak lensing mass calibration.  It does,
however, impact the uncertainties in concentration.  In short, current miscentring estimates are sufficiently
accurate to be negligible for mass calibration purposes, but not for analyses of the mass--concentration relation.

Since we have included our systematic uncertainties in the outputs of the MCMC chains themselves,
we cannot distinguish systematic from statistical errors in this calculation.  To obtain a statistical
error, we run a separate chain with all the nuisance parameters ($b$, the miscentring parameters,
and our bias correction) fixed to their central values, with no uncertainty included,
and measure the statistical error from the uncertainty in the resulting parameters.  
We subtract this in quadrature from the total systematic plus statistical 
uncertainty to obtain the systematic uncertainty.

\section{Results}\label{results}

\begin{figure*}
	\includegraphics[width=0.95\textwidth]{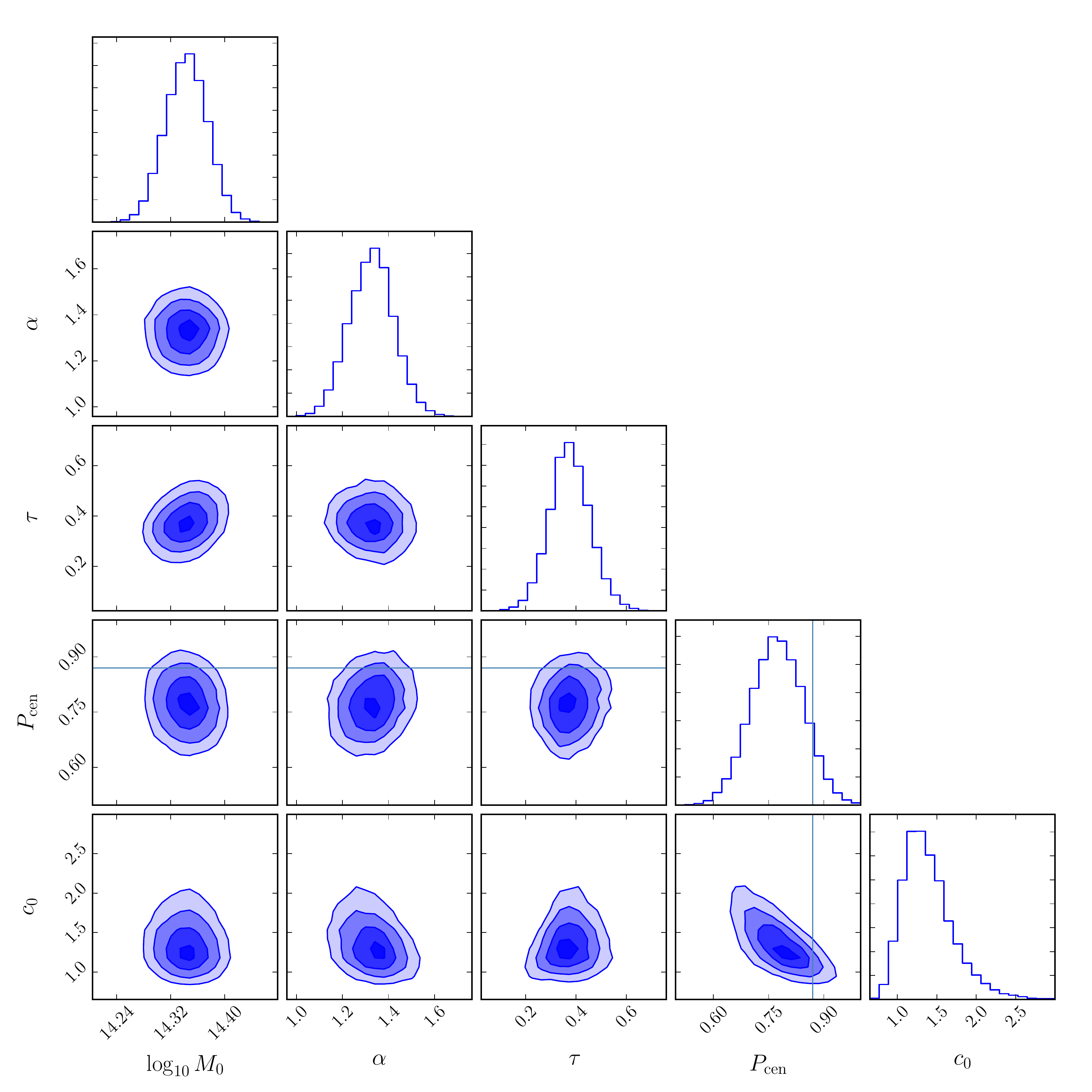}
	\caption{Results from our MCMC model fitting.  
		Contours indicate the 0.5, 1, 1.5, and 2$\sigma$ levels, and the 1 dimensional histograms are shown 
		on the diagonal.  The range shown for each parameter does not necessarily correspond
		to the top-hat prior ranges.  Parameters and priors are described in
		Table~\ref{params}.  The solid line denotes the expected value of $\left <
		P_{\mathrm{cen}} \right > = 0.87$ from the \redmapper{} catalogue.}\label{triangle-misvarc}
\end{figure*}

\begin{figure*}
\includegraphics[width=0.95\textwidth]{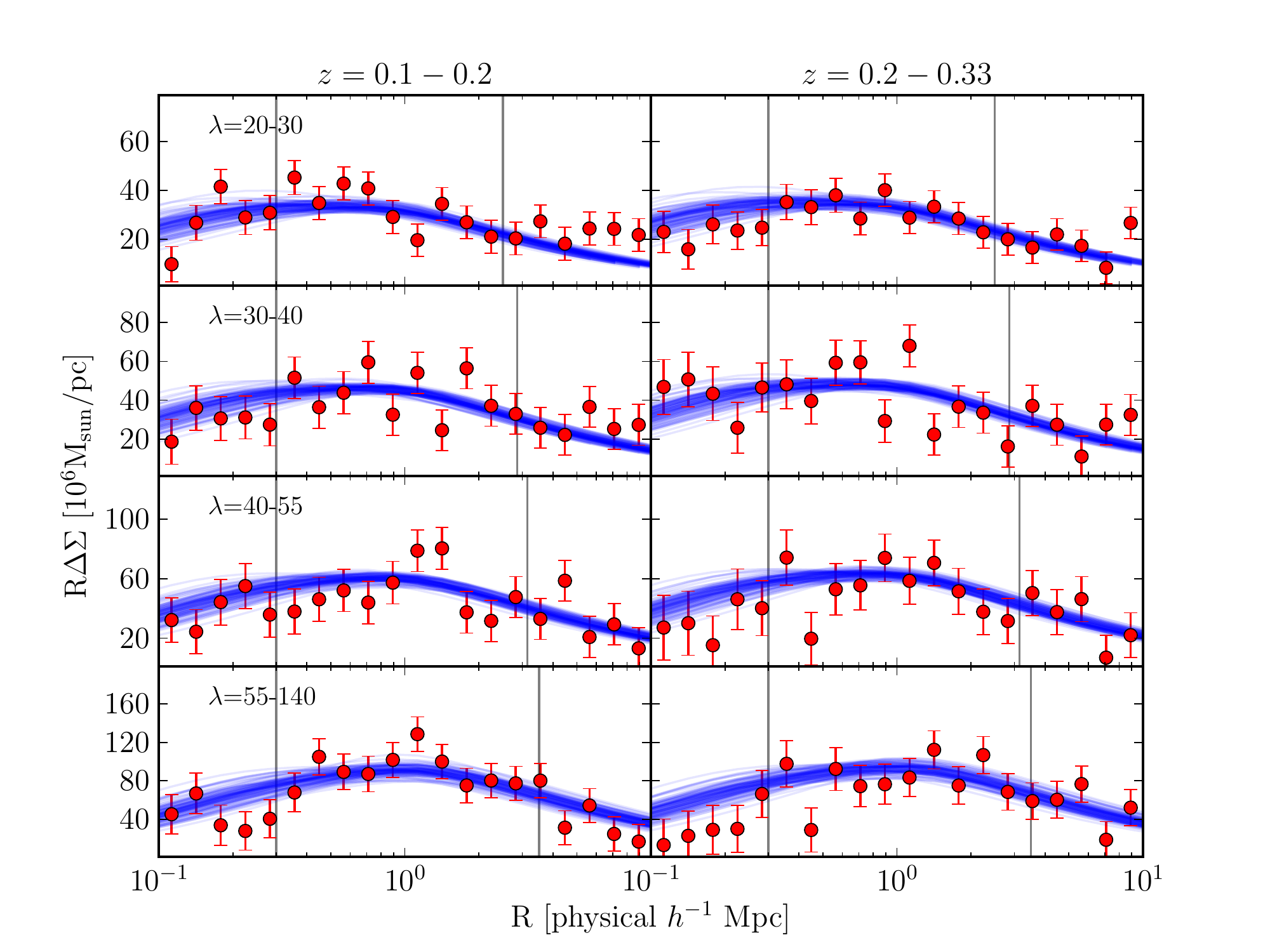}
\caption{Lensing signal ($R\Delta\Sigma$) from the \redmapper{} catalogue for the eight bins described in 
Table~\ref{bintable}, plus a sampling of models from the MCMC chain.  The fit was performed
only to data points lying between the vertical lines.}
\label{ds-misvarc}
\end{figure*}

\begin{figure}
\includegraphics[width=0.45\textwidth]{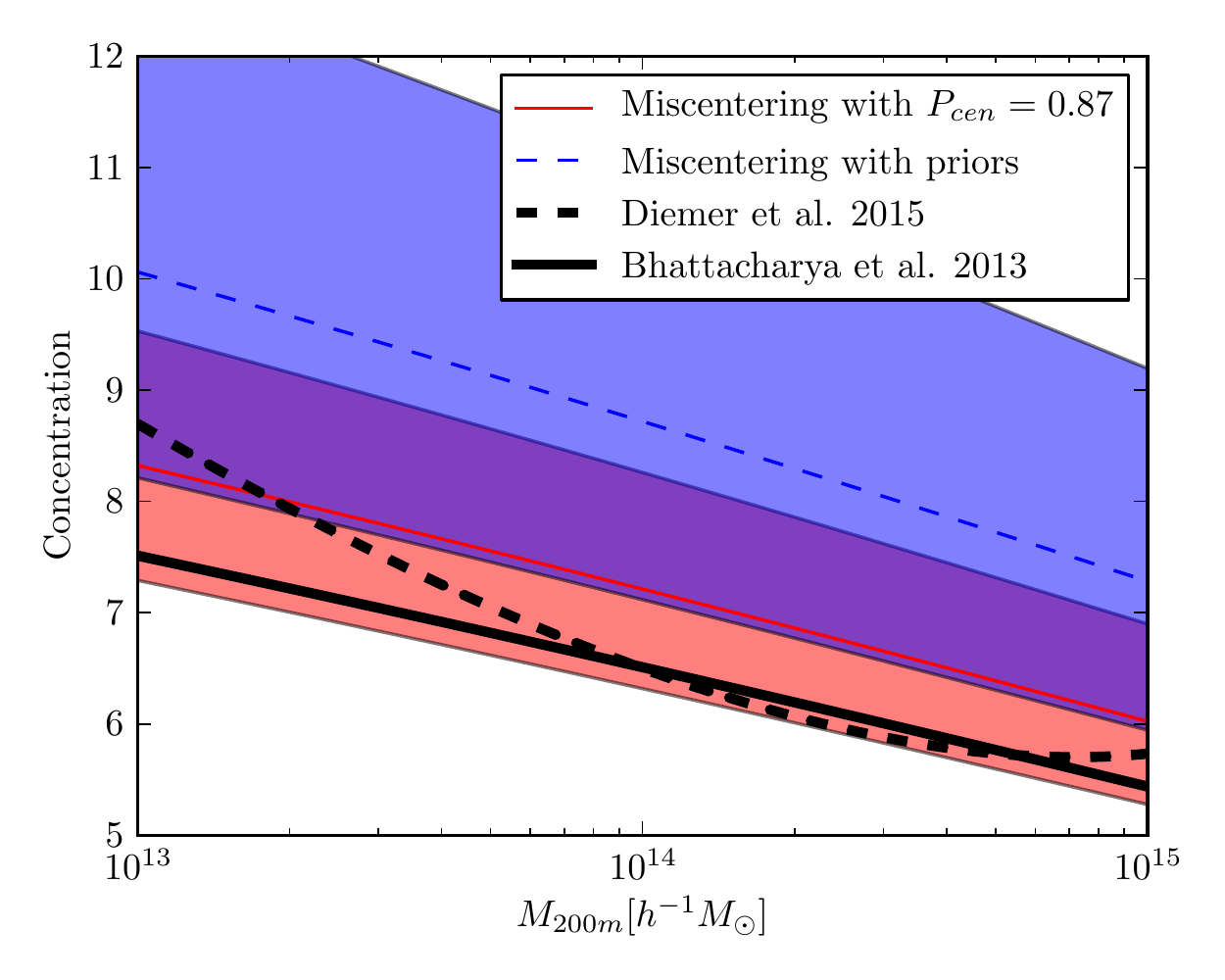}
\caption{The mass--concentration relations of \citet{2013ApJ...766...32B} and
\citet{2015ApJ...799..108D}, both at redshift $z=0.2$, along with the $1\sigma$ confidence
intervals for our $M-c$ relation with fitted amplitude, both with free miscentring and with
$\Pcen$ fixed to its value from the \redmapper\ catalogue.}\label{mc-comp}
\end{figure}

Figure~\ref{triangle-misvarc} shows the results from our MCMC fitting procedure.  We obtain a fit with $-2 \ln \lkhd=76.37$, which would
(without our priors) correspond to a $\chi^2$ value for 72 degrees of freedom. That is, our model is a good fit to the data.
A sampling of the models from our chains along with our data can be seen
in Figure~\ref{ds-misvarc}.  Table~\ref{params} reports our model parameters, along with the corresponding priors and posteriors.
We note that the amplitude and slope of the mass--richness scaling relation, as well as the amplitude of the mass--concentration relation,
are tightly constrained. 
Our constraint on the amplitude of mass--richness relation parameter is 
$\log M_0 = 14.344 \pm 0.021$ (statistical) $\pm 0.023$ (systematic),
corresponding to a $7\%$ calibration of the amplitude of the scaling relation. 
The corresponding constraint on
the power-law index of the mass-$\lambda$ relation is $\alpha = 1.33^{+0.09}_{-0.10}$.
The posterior probability for the scatter in the mass--richness relation and the miscentring
probability is largely unchanged from the input priors, demonstrating that
the external data sets utilized to derive the priors are significantly more sensitive to these parameters than the stacked weak lensing
signal measured in this work.  

There is a degeneracy between $\Pcen$ and the amplitude of the mass--concentration relation, which
is unsurprising, as both parameters correspond to shifts in the ``peakiness'' of the profile.  $\tau$ is not included in this
degeneracy, but shifts in $\tau$ have a smaller effect on the profile since $\Pcen$ is large.

We do not show the parameters $\sigma_{\ln M|\lambda}$ or $b$, which are unconstrained by the data.  $\sigma_{\ln M|\lambda}$ is not 
degenerate with any other parameter, and we find no preference for any value within our tophat prior range.  We also find no
preference for any value of $b$. Not surprisingly, we find that $b$ is degenerate with the mass, with the mass scaling as $10^{-b/2}$
for our fits. Also, $b$ is degenerate with the concentration amplitude $c_0$, with $c_0 \propto 1-b$.

\begin{table*}
\centering
\begin{tabular}{l l c c c}
\hline
Parameter & {\centering Description} & Prior & Median and error & Median and error \\
& & & (posterior) & ($\Pcen=0.87$) \\
\multirow{2}{*}{$\log_{10} M_0$} & \footnotesize{Log amplitude of scaling relation at $\lambda_0=40$} & \multirow{2}{*}{(13, 15)} & \multirow{2}{*}{ $14.344 \pm 0.021$ $\pm 0.023$} & \multirow{2}{*}{$14.338 \pm 0.021 \pm 0.021$} \\
& \footnotesize{in units of $h^{-1} M_{\odot}$, with definition $M_{200m}$}  & & & \\
$\alpha$ & \footnotesize{Power-law index of dependence on $\lambda$} & (0, 2) & $1.33^{+0.09}_{-0.10}$  & $1.34 \pm 0.09$ \\
$\sigma_{\ln M|\lambda}$ & \footnotesize{Intrinsic scatter of $\lambda-M$ relation} & (0.2, 0.3) & $0.25 \pm 0.03\dagger$  & $0.25 \pm 0.03\dagger$ \\
\multirow{2}{*}{$c_0$} & \footnotesize{Amplitude of mass--concentration relation} & \multirow{2}{*}{[0, 3)} &  \multirow{2}{*}{$1.34^{+0.35}_{-0.25}$} & \multirow{2}{*}{$1.11^{+0.16}_{-0.14}$} \\
 & \footnotesize{as a multiple of \citet{2013ApJ...766...32B}} & &  & \\
$\tau$ & \footnotesize{Width of the Gaussian miscentring distribution} & $\mathcal{N}(0.4,0.1)$ & $0.37 \pm 0.08\dagger$  & $0.38 \pm 0.09\dagger$ \\
$\Pcen$ & \footnotesize{Fraction of clusters with correct centres} & $\mathcal{N}(0.80,0.07)$ & $0.77 \pm 0.07\dagger$  & --- \\
\multirow{2}{*}{$b$} & \footnotesize{Multiplicative amplitude shift to accommodate} & \multirow{2}{*}{(-0.05, 0.05)} & \multirow{2}{*}{$0.00 \pm 0.03\dagger$}  & \multirow{2}{*}{$0.00 \pm 0.03\dagger$} \\
& \footnotesize{systematic errors in $\Delta\Sigma$} & & &  \\
\end{tabular} \\
\caption{
Parameters from Eqs.~\eqref{massmodel} and~\eqref{deltasigmamodel}, a short description of the 
meaning, the prior, and the median of all MCMC samples plus the errors given by the 16th and 84th percentiles
of the samples.  $M_0$ has been corrected from the output of the MCMC chain for model bias, as described
in Section~\ref{sec:sys}.  We also show the results if we fix
$\Pcen$ to 0.87, the mean value from the \redmapper\ catalogue centring probabilities.
A range means a tophat prior; $\mathcal{N}(x,\sigma)$ is a Gaussian prior with mean $x$ and 
width $\sigma$.  For posteriors with two uncertainties given, the first is statistical and
the second is systematic.
The $\dagger$ symbol indicates values that are largely determined by our priors rather than our data.
}\label{params}
\end{table*}

In Fig.~\ref{mc-comp}, we show how our fitted mass--concentration relation compares to two theoretical models,
\citet{2013ApJ...766...32B} and \citet{2015ApJ...799..108D}.  We also show the mass--concentration relation recovered
when we 
fix $\Pcen$ to the value expected from the \redmapper\ catalogue, 0.87.  In both cases, 
we find somewhat higher values of $c$
than the theoretical models predict, but the difference is only $1\sigma$ and hence not
statistically significant.
Appendix~\ref{app:redshift} also demonstrates that allowing for redshift evolution
in the mass--richness relation does not alter any of our conclusions.  In Appendix~\ref{app:models} we show results with and
without various complicating factors (such as miscentring and a variable mass--concentration 
relation), to aid comparison with other works and to show the
effects of those factors for interested readers.

Because the halo mass definition is dependent on cosmology, our results change if we alter $\Omega_m$.  We perform the same
analysis as above with $\Omega_m$ from $0.26-0.34$ to check the dependence of our parameters on cosmology, altering
$\Omega_{\Lambda}$ as well to maintain a flat universe.  Only the
mass amplitude changes, as expected; we find that $\log_{10} M_0$ is linear with $\Omega_m$ near $\Omega_m=0.3$, with
the scaling $\log_{10} M_0 = 14.344 - 0.706(\Omega_m-0.3)$.  As expected, $M_0$ decreases as $\Omega_m$ increases.

\section{Discussion}\label{conclusions}

We compare our results to a variety of constraints from the literature.
The first mass--richness relation for \redmapper\ clusters was a rough abundance-matching
estimate presented in \citet{rykoffetal12}, who suggested a scaling relation with $\alpha=1.08$
and $M=(3.9 \pm 1.2) \times 10^{14}\ \hMsun$ at $\lambda=60$, in excellent
agreement with our recovered value of  $(3.7 \pm 0.3) \times 10^{14}\ \hMsun$.  There is 
an apparent slight tension ($2.5\sigma$) in the value of the slope, but \citet{rykoffetal12} did not
report an uncertainty in the slope.  We note that \citet{rykoffetal12} emphasized their reported
relation was only meant to be a place holder for subsequent mass calibration efforts such as this one.

\citet{2015arXiv150606135M} used the same SDSS weak 
lensing shear catalogue and \photoz catalogue and calibration as we did to constrain the mean mass of
entire \redmapper\ cluster catalogue.  Despite the shared input data set, there are significant methodological 
differences between the two analyses.  In particular, our science goal led us to provide a much more detailed
systematics analysis than that presented in \citet{2015arXiv150606135M}.  The end mass calibrations are,
perhaps not surprisingly, very similar.  Our principal contribution is the quantitative
characterization of the systematic uncertainties inherent to this measurement, as well as
determination of the scaling of mass with $\lambda$ instead of just a population average.

\citet{lietal15} produced a weak lensing calibration of the mass--richness relation as a by-product of their analysis on 
the lensing profile of cluster substructures.  Specifically, they assume a mass--richness relation that
was identical to that of \citet{rykoffetal12}, except modulated by an amplitude $A$.  This amplitude
is allowed to float when considering substructures in different radial bins. We combine the various
radial bins, ignoring the innermost bin because it is discrepant with the rest of the data, and is the
bin most likely to be affected by systematic uncertainties from magnification, non-weak shear, and
source obscuration.  Averaging the \citet{lietal15} data results in $A=0.803 \pm 0.016$, corresponding to 
$M=(2.09\pm 0.04)\times 10^{14}\ \hMsun$ at $\lambda=40$.
This value is in excellent agreement
with our recovered value of $M=(2.21\pm 0.15)\times 10^{14}\ \hMsun$.   We note the error quoted
in \citet{lietal15} is statistical only.
 
Recent work by \citet{farahietal16} has calibrated the mass--richness relation of \redmapper\ clusters
using stacked velocity dispersion information.  They report $M_{200c}=(1.56 \pm 0.35) \times 10^{14}\ \msun$
at $\lambda=30$ with $h=0.7$.  This can be compared to our prediction
$M_{200c}=  (1.60\pm 0.11) \times 10^{14}\ \msun$,
in excellent agreement with their result.  The corresponding slopes are also in excellent agreement,
$\alpha=1.33^{+0.09}_{-0.10}$ in our analysis, and $\alpha=1.31\pm 0.14$ in theirs.

\citet[][hereafter referred to simply as Saro]{2015MNRAS.454.2305S} calibrated the richness--mass relation 
for \redmapper\ clusters in
the Dark Energy Survey Science (DES) Verification data by cross-matching DES \redmapper\ clusters
to clusters found by the South Pole Telescope \citep[SPT][]{2015ApJS..216...27B}.  The SPT clusters were 
assigned masses by assuming the SPT cluster abundance is consistent with a flat $\Lambda$CDM
cosmology with $\Omega_m=0.3$ and $\sigma_8=0.8$.  In this way, the SPT clusters are assigned
both a mass --- for which they utilized $M_{500c}$ --- and a richness, 
allowing \citet{2015MNRAS.454.2305S} to infer the cluster richness--mass
relation.  Using the framework of \citet{2014MNRAS.441.3562E}, the corresponding mass--richness
relation can be written as $\avg{M_{500c}|\lambda} = (3.2\pm 0.6)\times 10^{14}\ \hMsun$ at $\lambda=66.1$,
the pivot richness for their sample.  The above value is corrected for the predicted redshift evolution between
the Saro redshift pivot point and ours, using the Saro constraints on the evolution of the mass--richness relation,
and the corresponding uncertainty has been adequately propagated.
We convert our predicted masses from $M_{200m}$ to $M_{500c}$
following \citet{hukravtsov03} and the \citet{2013ApJ...766...32B} mass--concentration relation
to arrive at a predicted mass $M_{500c}=(2.2\pm 0.15)\times 10^{14}\ \hMsun$, in reasonable 
agreement ($1.6\sigma$)
with the Saro value.  Despite the apparent disparity in the central values for the slopes of the mass--richness relation ---
Saro finds $\alpha=0.91 \pm 0.18$, compared to our value
$1.33^{+0.09}_{-0.10}$ --- the two values are consistent at the $2\sigma$ level. 
We note that the above comparison assumes there is no systematic offset in the richness
measurements of a galaxy cluster between the SDSS and DES data sets, an assumption
that is currently not testable.  Indeed, the agreement between the two measurements could
be interpreted as evidence that there are no large systematic differences in the richness between
the two data sets, as suggested by a comparison of their relative abundances \citep{2016arXiv160100621R}.

Our miscentring values are largely controlled by our priors, as we do not have as much constraining
power with the weak lensing only due to degeneracies between parameters that modify the profile shape.
We can compare these miscentring parameters to previous results as well.  Our lensing-weighted average $R_{\lambda}$ is 0.79, so our
miscentring $\tau$ corresponds to approximately $R_{\textrm{mis}}=0.29\pm 0.06h^{-1}$ Mpc.  
\citet{rozorykoff14} find offsets corresponding to 
constant probability out to $R_{\textrm{mis}}=0.8 h^{-1}$~Mpc.  However, we note that their straight-line fit is in fact similar to 
the cumulative distribution function of the Rayleigh distribution we
assume for our miscentring offsets. The constant probability out to
$R_{\textrm{mis}}=0.8h^{-1}$ Mpc that they find is approximately equivalent to a Rayleigh distribution with 
$R_{\textrm{mis}}=0.3h^{-1}$ Mpc, very consistent with our results.  Our peak value is 
smaller than the simulations of the MaxBCG cluster sample \citep{2007arXiv0709.1159J} of $\sim 0.4h^{-1}$ Mpc,
although as that catalogue uses different methods of finding the centre of a galaxy clusters, 
discrepancies should be expected.  \red{The RCS cluster sample has a well-measured miscentering width of
$0.41 \pm 0.01 h^{-1}$ Mpc for $10^{14} \msun$ halos \citep{2016A&A...586A..43V}, in good agreement with our measurement of
$0.43 \pm 0.09 h^{-1}$ Mpc for the same halos. Their fraction of correctly-centred halos is smaller,
again reflecting differences in the cluster-finding algorithms or perhaps differences in the performance
of the centroiding algorithm with redshift, as they include clusters up to $z=0.7$.}

As for the mass--concentration relation, our results appear to be in excellent agreement with theoretical
expectations.  We caution, however, that we while our model account for projection effects, halo triaxiality, 
and cluster miscentring with regards to calibrating the mass--richness relation, we have not performed
a similar calibration for the impact on concentration.  Indeed, a recent paper by Baxter et al (in prep), looking
at the clustering of \redmapper\ clusters, demonstrated that \redmapper\ clusters tend to be preferentially
more compact than randomly oriented halos, an effect which would tend to bias our recovered parameter
$c_0$ to values larger than unity, exactly as observed.   We postpone a detailed calibration of this
effect, and therefore a more detailed comparison to numerical simulations and observational studies, to future work.

\section{Summary}

We have measured the weak lensing signal around 5,570 clusters in the \redmapper\ catalogue from SDSS DR8, with richnesses
$20<\lambda<140$ and redshifts $0.1<z<0.33$.  The signal shows good agreement with a comparison signal calculated using the
same cluster catalogue, but a different shape measurement pipeline and \photoz{} algorithm and pipeline.  

The mass modeling method we use is an extension of previous work, such as the maximum-likelihood lensing approach adopted by
\citet{2015MNRAS.446.1356H} and the combination fits of \citet{2014MNRAS.439.3755F}.  The model
makes it easy to self-consistently include a varying mass--concentration
relation and scatter in the mass--richness and mass--concentration relations, as well as removing uncertainty from the question
of which richness in a richness bin best corresponds to the mass value obtained from a single-halo
fit. We test the model
on $N$-body simulations and find excellent agreement with the true input cluster sample properties.

We fit a parameterized model for the mass--richness relation to the lensing signal, allowing the
mass--concentration relation to
float to obtain a good fit, accounting for cluster miscentring and scatter in the mass--richness relation, and marginalizing over remaining multiplicative
errors in our lensing measurement. We find a mass--richness relation of
\begin{equation}
\avg{M_{200m}|\lambda} = 10^{14.344 \pm 0.021\ \mathrm{stat.}\pm 0.023\ \mathrm{sys.}} \left(\frac{\lambda}{40}\right)^{1.33^{+0.09}_{-0.10}} 
\end{equation}
which is consistent with previous measurements using lensing, 
and in modest agreement with results from the SZ effect.

Including miscentring and a varying mass--concentration relation are important to obtaining accurate results with this sample in this
radius range.
That is, while our priors on cluster centring are sufficiently tight that miscentring does not contribute a significant fraction of the variance in our mass calibration, we find that ignoring centring outright would result in a $0.3\sigma$ systematic
shift in the slope of our mass--richness relation towards higher values.

The results in this work provide the most careful weak lensing mass calibration analysis of the \redmapper\
cluster catalogue to date, with a detailed budget of systematic uncertainties and null tests.  This is also
the first time that a cluster mass calibration effort has included a null test comparing two independently developed
\photoz\ and shear codes as a way to validate the estimated systematic uncertainties in the recovered
halo masses.  Our results provide a critical stepping stone towards placing cosmological constraints
with the \redmapper\ cluster sample, and pave the way for similar analyses in upcoming photometric
survey data, such as that of the DES, HSC, and LSST surveys.

\section*{Acknowledgements}

The authors thank Bhuvnesh Jain and Eric Baxter for useful discussions
related to this work.  MS and RM acknowledge the support of the
Department of Energy Early Career Award program.  ES is supported by
DOE grant DE-AC02-98CH10886.  This work received partial support from
the U.S.\ Department of Energy under contract number DE-AC02-76SF00515
and from the National Science Foundation under NSF-AST-1211838.  We
thank Mathew Becker and Michael Busha for their contributions to the
N-body simulations used in this work.

Funding for SDSS-III has been provided by the Alfred P. Sloan Foundation, the Participating Institutions, the National Science Foundation, and the U.S. Department of Energy Office of Science. The SDSS-III web site is http://www.sdss3.org/.

SDSS-III is managed by the Astrophysical Research Consortium for the Participating Institutions of the SDSS-III Collaboration including the University of Arizona, the Brazilian Participation Group, Brookhaven National Laboratory, Carnegie Mellon University, University of Florida, the French Participation Group, the German Participation Group, Harvard University, the Instituto de Astrofisica de Canarias, the Michigan State/Notre Dame/JINA Participation Group, Johns Hopkins University, Lawrence Berkeley National Laboratory, Max Planck Institute for Astrophysics, Max Planck Institute for Extraterrestrial Physics, New Mexico State University, New York University, Ohio State University, Pennsylvania State University, University of Portsmouth, Princeton University, the Spanish Participation Group, University of Tokyo, University of Utah, Vanderbilt University, University of Virginia, University of Washington, and Yale University. 

\bibliographystyle{mnras}
\bibliography{redmapper_wl}

\begin{thebibliography}{}
\makeatletter
\relax
\def\mn@urlcharsother{\let\do\@makeother \do\$\do\&\do\#\do\^\do\_\do\%\do\~}
\def\mn@doi{\begingroup\mn@urlcharsother \@ifnextchar [ {\mn@doi@}
  {\mn@doi@[]}}
\def\mn@doi@[#1]#2{\def\@tempa{#1}\ifx\@tempa\@empty \href
  {http://dx.doi.org/#2} {doi:#2}\else \href {http://dx.doi.org/#2} {#1}\fi
  \endgroup}
\def\mn@eprint#1#2{\mn@eprint@#1:#2::\@nil}
\def\mn@eprint@arXiv#1{\href {http://arxiv.org/abs/#1} {{\tt arXiv:#1}}}
\def\mn@eprint@dblp#1{\href {http://dblp.uni-trier.de/rec/bibtex/#1.xml}
  {dblp:#1}}
\def\mn@eprint@#1:#2:#3:#4\@nil{\def\@tempa {#1}\def\@tempb {#2}\def\@tempc
  {#3}\ifx \@tempc \@empty \let \@tempc \@tempb \let \@tempb \@tempa \fi \ifx
  \@tempb \@empty \def\@tempb {arXiv}\fi \@ifundefined
  {mn@eprint@\@tempb}{\@tempb:\@tempc}{\expandafter \expandafter \csname
  mn@eprint@\@tempb\endcsname \expandafter{\@tempc}}}

\bibitem[\protect\citeauthoryear{{Abazajian} \& {Dodelson}}{{Abazajian} \&
  {Dodelson}}{2003}]{2003PhRvL..91d1301A}
{Abazajian} K.,  {Dodelson} S.,  2003, \mn@doi [Physical Review Letters]
  {10.1103/PhysRevLett.91.041301}, \href
  {http://adsabs.harvard.edu/abs/2003PhRvL..91d1301A} {91, 041301}

\bibitem[\protect\citeauthoryear{{Aihara} et~al.,}{{Aihara}
  et~al.}{2011}]{2011ApJS..193...29A}
{Aihara} H.,  et~al., 2011, \mn@doi [\apjs] {10.1088/0067-0049/193/2/29}, \href
  {http://adsabs.harvard.edu/abs/2011ApJS..193...29A} {193, 29}

\bibitem[\protect\citeauthoryear{{Albrecht} et~al.,}{{Albrecht}
  et~al.}{2006}]{DarkEnrgTF}
{Albrecht} A.,  et~al., 2006, ArXiv Astrophysics e-prints, \href
  {http://adsabs.harvard.edu/abs/2006astro.ph..9591A} {}

\bibitem[\protect\citeauthoryear{{Allen}, {Evrard}  \& {Mantz}}{{Allen}
  et~al.}{2011}]{2011ARAA..49..409A}
{Allen} S.~W.,  {Evrard} A.~E.,   {Mantz} A.~B.,  2011, \mn@doi [\araa]
  {10.1146/annurev-astro-081710-102514}, \href
  {http://adsabs.harvard.edu/abs/2011ARA\%26A..49..409A} {49, 409}

\bibitem[\protect\citeauthoryear{{Angulo}, {Springel}, {White}, {Jenkins},
  {Baugh}  \& {Frenk}}{{Angulo} et~al.}{2012}]{anguloetal12}
{Angulo} R.~E.,  {Springel} V.,  {White} S.~D.~M.,  {Jenkins} A.,  {Baugh}
  C.~M.,   {Frenk} C.~S.,  2012, \mn@doi [\mnras]
  {10.1111/j.1365-2966.2012.21830.x}, \href
  {http://adsabs.harvard.edu/abs/2012MNRAS.426.2046A} {426, 2046}

\bibitem[\protect\citeauthoryear{{Bacon}, {Refregier}  \& {Ellis}}{{Bacon}
  et~al.}{2000}]{2000MNRAS.318..625B}
{Bacon} D.~J.,  {Refregier} A.~R.,   {Ellis} R.~S.,  2000, \mn@doi [\mnras]
  {10.1046/j.1365-8711.2000.03851.x}, \href
  {http://adsabs.harvard.edu/abs/2000MNRAS.318..625B} {318, 625}

\bibitem[\protect\citeauthoryear{{Bartelmann}}{{Bartelmann}}{1996}]{BartelmannNFW1996}
{Bartelmann} M.,  1996, \aap, \href
  {http://adsabs.harvard.edu/abs/1996A%26A...313..697B} {313, 697}

\bibitem[\protect\citeauthoryear{{Bartelmann} \& {Schneider}}{{Bartelmann} \&
  {Schneider}}{2001}]{2001PhR...340..291B}
{Bartelmann} M.,  {Schneider} P.,  2001, \mn@doi [\physrep]
  {10.1016/S0370-1573(00)00082-X}, \href
  {http://adsabs.harvard.edu/abs/2001PhR...340..291B} {340, 291}

\bibitem[\protect\citeauthoryear{{Becker} \& {Kravtsov}}{{Becker} \&
  {Kravtsov}}{2011}]{beckerkravtsov11}
{Becker} M.~R.,  {Kravtsov} A.~V.,  2011, \mn@doi [\apj]
  {10.1088/0004-637X/740/1/25}, \href
  {http://adsabs.harvard.edu/abs/2011ApJ...740...25B} {740, 25}

\bibitem[\protect\citeauthoryear{{Becker} et~al.,}{{Becker}
  et~al.}{2016}]{2015arXiv150705598B}
{Becker} M.~R.,  et~al., 2016, \mn@doi [\prd] {10.1103/PhysRevD.94.022002},
  \href {http://adsabs.harvard.edu/abs/2016PhRvD..94b2002B} {94, 022002}

\bibitem[\protect\citeauthoryear{{Behroozi}, {Wechsler}  \& {Wu}}{{Behroozi}
  et~al.}{2013}]{Behroozi2013a}
{Behroozi} P.~S.,  {Wechsler} R.~H.,   {Wu} H.-Y.,  2013, \mn@doi [\apj]
  {10.1088/0004-637X/762/2/109}, \href
  {http://adsabs.harvard.edu/abs/2013ApJ...762..109B} {762, 109}

\bibitem[\protect\citeauthoryear{{Ben{\'{\i}}tez}}{{Ben{\'{\i}}tez}}{2000}]{2000ApJ...536..571B}
{Ben{\'{\i}}tez} N.,  2000, \mn@doi [ApJ] {10.1086/308947}, \href
  {http://adsabs.harvard.edu/abs/2000ApJ...536..571B} {536, 571}

\bibitem[\protect\citeauthoryear{{Bernstein} \& {Jarvis}}{{Bernstein} \&
  {Jarvis}}{2002}]{2002AJ....123..583B}
{Bernstein} G.~M.,  {Jarvis} M.,  2002, \mn@doi [AJ] {10.1086/338085}, \href
  {http://adsabs.harvard.edu/abs/2002AJ....123..583B} {123, 583}

\bibitem[\protect\citeauthoryear{{Bhattacharya}, {Habib}, {Heitmann}  \&
  {Vikhlinin}}{{Bhattacharya} et~al.}{2013}]{2013ApJ...766...32B}
{Bhattacharya} S.,  {Habib} S.,  {Heitmann} K.,   {Vikhlinin} A.,  2013,
  \mn@doi [\apj] {10.1088/0004-637X/766/1/32}, \href
  {http://adsabs.harvard.edu/abs/2013ApJ...766...32B} {766, 32}

\bibitem[\protect\citeauthoryear{{Blazek}, {Mandelbaum}, {Seljak}  \&
  {Nakajima}}{{Blazek} et~al.}{2012}]{2012JCAP...05..041B}
{Blazek} J.,  {Mandelbaum} R.,  {Seljak} U.,   {Nakajima} R.,  2012, \mn@doi
  [\jcap] {10.1088/1475-7516/2012/05/041}, \href
  {http://adsabs.harvard.edu/abs/2012JCAP...05..041B} {5, 041}

\bibitem[\protect\citeauthoryear{{Bleem} et~al.,}{{Bleem}
  et~al.}{2015}]{2015ApJS..216...27B}
{Bleem} L.~E.,  et~al., 2015, \mn@doi [\apjs] {10.1088/0067-0049/216/2/27},
  \href {http://adsabs.harvard.edu/abs/2015ApJS..216...27B} {216, 27}

\bibitem[\protect\citeauthoryear{{Bocquet}, {Saro}, {Dolag}  \&
  {Mohr}}{{Bocquet} et~al.}{2016}]{bocquet16}
{Bocquet} S.,  {Saro} A.,  {Dolag} K.,   {Mohr} J.~J.,  2016, \mn@doi [\mnras]
  {10.1093/mnras/stv2657}, \href
  {http://adsabs.harvard.edu/abs/2016MNRAS.456.2361B} {456, 2361}

\bibitem[\protect\citeauthoryear{{Brainerd}, {Blandford}  \&
  {Smail}}{{Brainerd} et~al.}{1996}]{1996ApJ...466..623B}
{Brainerd} T.~G.,  {Blandford} R.~D.,   {Smail} I.,  1996, \mn@doi [\apj]
  {10.1086/177537}, \href {http://adsabs.harvard.edu/abs/1996ApJ...466..623B}
  {466, 623}

\bibitem[\protect\citeauthoryear{{Broadhurst}, {Taylor}  \&
  {Peacock}}{{Broadhurst} et~al.}{1995}]{broadhurstetal95}
{Broadhurst} T.~J.,  {Taylor} A.~N.,   {Peacock} J.~A.,  1995, \mn@doi [\apj]
  {10.1086/175053}, \href {http://adsabs.harvard.edu/abs/1995ApJ...438...49B}
  {438, 49}

\bibitem[\protect\citeauthoryear{{Cavagnolo}, {Donahue}, {Voit}  \&
  {Sun}}{{Cavagnolo} et~al.}{2009}]{2009ApJS..182...12C}
{Cavagnolo} K.~W.,  {Donahue} M.,  {Voit} G.~M.,   {Sun} M.,  2009, \mn@doi
  [\apjs] {10.1088/0067-0049/182/1/12}, \href
  {http://adsabs.harvard.edu/abs/2009ApJS..182...12C} {182, 12}

\bibitem[\protect\citeauthoryear{{Clampitt} \& {Jain}}{{Clampitt} \&
  {Jain}}{2015}]{2014ClampittVoids}
{Clampitt} J.,  {Jain} B.,  2015, \mn@doi [\mnras] {10.1093/mnras/stv2215},
  \href {http://adsabs.harvard.edu/abs/2015MNRAS.454.3357C} {454, 3357}

\bibitem[\protect\citeauthoryear{{Corless} \& {King}}{{Corless} \&
  {King}}{2009}]{2009MNRAS.396..315C}
{Corless} V.~L.,  {King} L.~J.,  2009, \mn@doi [\mnras]
  {10.1111/j.1365-2966.2009.14542.x}, \href
  {http://adsabs.harvard.edu/abs/2009MNRAS.396..315C} {396, 315}

\bibitem[\protect\citeauthoryear{{Coupon}, {Broadhurst}  \& {Umetsu}}{{Coupon}
  et~al.}{2013}]{couponetal13}
{Coupon} J.,  {Broadhurst} T.,   {Umetsu} K.,  2013, \mn@doi [\apj]
  {10.1088/0004-637X/772/1/65}, \href
  {http://adsabs.harvard.edu/abs/2013ApJ...772...65C} {772, 65}

\bibitem[\protect\citeauthoryear{{Cui} et~al.,}{{Cui}
  et~al.}{2016a}]{2016MNRAS.456.2566C}
{Cui} W.,  et~al., 2016a, \mn@doi [\mnras] {10.1093/mnras/stv2839}, \href
  {http://adsabs.harvard.edu/abs/2016MNRAS.456.2566C} {456, 2566}

\bibitem[\protect\citeauthoryear{{Cui} et~al.,}{{Cui}
  et~al.}{2016b}]{2016arXiv160206668C}
{Cui} W.,  et~al., 2016b, \mn@doi [\mnras] {10.1093/mnras/stw603}, \href
  {http://adsabs.harvard.edu/abs/2016MNRAS.458.4052C} {458, 4052}

\bibitem[\protect\citeauthoryear{{Dawson} et~al.,}{{Dawson}
  et~al.}{2013}]{2013AJ....145...10D}
{Dawson} K.~S.,  et~al., 2013, \mn@doi [\aj] {10.1088/0004-6256/145/1/10},
  \href {http://adsabs.harvard.edu/abs/2013AJ....145...10D} {145, 10}

\bibitem[\protect\citeauthoryear{{Diemer} \& {Kravtsov}}{{Diemer} \&
  {Kravtsov}}{2014}]{2014ApJ...789....1D}
{Diemer} B.,  {Kravtsov} A.~V.,  2014, \mn@doi [\apj]
  {10.1088/0004-637X/789/1/1}, \href
  {http://adsabs.harvard.edu/abs/2014ApJ...789....1D} {789, 1}

\bibitem[\protect\citeauthoryear{{Diemer} \& {Kravtsov}}{{Diemer} \&
  {Kravtsov}}{2015}]{2015ApJ...799..108D}
{Diemer} B.,  {Kravtsov} A.~V.,  2015, \mn@doi [\apj]
  {10.1088/0004-637X/799/1/108}, \href
  {http://adsabs.harvard.edu/abs/2015ApJ...799..108D} {799, 108}

\bibitem[\protect\citeauthoryear{{Dietrich} et~al.,}{{Dietrich}
  et~al.}{2014}]{dietrichetal14}
{Dietrich} J.~P.,  et~al., 2014, \mn@doi [\mnras] {10.1093/mnras/stu1282},
  \href {http://adsabs.harvard.edu/abs/2014MNRAS.443.1713D} {443, 1713}

\bibitem[\protect\citeauthoryear{{Du}, {Fan}, {Shan}, {Zhao}, {Covone}, {Fu}
  \& {Kneib}}{{Du} et~al.}{2015}]{2015arXiv151008193D}
{Du} W.,  {Fan} Z.,  {Shan} H.,  {Zhao} G.-B.,  {Covone} G.,  {Fu} L.,
  {Kneib} J.-P.,  2015, \mn@doi [\apj] {10.1088/0004-637X/814/2/120}, \href
  {http://adsabs.harvard.edu/abs/2015ApJ...814..120D} {814, 120}

\bibitem[\protect\citeauthoryear{{Dutton} \& {Macci{\`o}}}{{Dutton} \&
  {Macci{\`o}}}{2014}]{duttonmaccio14}
{Dutton} A.~A.,  {Macci{\`o}} A.~V.,  2014, \mn@doi [\mnras]
  {10.1093/mnras/stu742}, \href
  {http://adsabs.harvard.edu/abs/2014MNRAS.441.3359D} {441, 3359}

\bibitem[\protect\citeauthoryear{Einasto}{Einasto}{1965}]{Einasto}
Einasto J.,  1965, Trudy Astrofizicheskogo Instituta Alma-Ata, 5, 87

\bibitem[\protect\citeauthoryear{{Eisenstein} et~al.,}{{Eisenstein}
  et~al.}{2011}]{2011AJ....142...72E}
{Eisenstein} D.~J.,  et~al., 2011, \mn@doi [\aj] {10.1088/0004-6256/142/3/72},
  \href {http://adsabs.harvard.edu/abs/2011AJ....142...72E} {142, 72}

\bibitem[\protect\citeauthoryear{{Evrard}, {Arnault}, {Huterer}  \&
  {Farahi}}{{Evrard} et~al.}{2014}]{2014MNRAS.441.3562E}
{Evrard} A.~E.,  {Arnault} P.,  {Huterer} D.,   {Farahi} A.,  2014, \mn@doi
  [\mnras] {10.1093/mnras/stu784}, \href
  {http://adsabs.harvard.edu/abs/2014MNRAS.441.3562E} {441, 3562}

\bibitem[\protect\citeauthoryear{{Farahi}, {Evrard}, {Rozo}, {Rykoff}  \&
  {Wechsler}}{{Farahi} et~al.}{2016}]{farahietal16}
{Farahi} A.,  {Evrard} A.~E.,  {Rozo} E.,  {Rykoff} E.~S.,   {Wechsler} R.~H.,
  2016, \mn@doi [\mnras] {10.1093/mnras/stw1143}, \href
  {http://adsabs.harvard.edu/abs/2016MNRAS.460.3900F} {460, 3900}

\bibitem[\protect\citeauthoryear{{Feldmann} et~al.}{{Feldmann}
  et~al.}{2006}]{2006MNRAS.372..565F}
{Feldmann} R.,  et~al., 2006, \mn@doi [MNRAS]
  {10.1111/j.1365-2966.2006.10930.x}, \href
  {http://adsabs.harvard.edu/abs/2006MNRAS.372..565F} {372, 565}

\bibitem[\protect\citeauthoryear{{Fischer} et~al.,}{{Fischer}
  et~al.}{2000}]{FischerGGL2000}
{Fischer} P.,  et~al., 2000, \mn@doi [\aj] {10.1086/301540}, \href
  {http://adsabs.harvard.edu/abs/2000AJ....120.1198F} {120, 1198}

\bibitem[\protect\citeauthoryear{{Ford}, {Hildebrandt}, {Van Waerbeke},
  {Erben}, {Laigle}, {Milkeraitis}  \& {Morrison}}{{Ford}
  et~al.}{2014}]{2014MNRAS.439.3755F}
{Ford} J.,  {Hildebrandt} H.,  {Van Waerbeke} L.,  {Erben} T.,  {Laigle} C.,
  {Milkeraitis} M.,   {Morrison} C.~B.,  2014, \mn@doi [\mnras]
  {10.1093/mnras/stu225}, \href
  {http://adsabs.harvard.edu/abs/2014MNRAS.439.3755F} {439, 3755}

\bibitem[\protect\citeauthoryear{{Foreman-Mackey}, {Hogg}, {Lang}  \&
  {Goodman}}{{Foreman-Mackey} et~al.}{2013}]{2013PASP..125..306F}
{Foreman-Mackey} D.,  {Hogg} D.~W.,  {Lang} D.,   {Goodman} J.,  2013, \mn@doi
  [\pasp] {10.1086/670067}, \href
  {http://adsabs.harvard.edu/abs/2013PASP..125..306F} {125, 306}

\bibitem[\protect\citeauthoryear{{Han} et~al.,}{{Han}
  et~al.}{2015}]{2015MNRAS.446.1356H}
{Han} J.,  et~al., 2015, \mn@doi [\mnras] {10.1093/mnras/stu2178}, \href
  {http://adsabs.harvard.edu/abs/2015MNRAS.446.1356H} {446, 1356}

\bibitem[\protect\citeauthoryear{{Hirata} \& {Seljak}}{{Hirata} \&
  {Seljak}}{2003}]{2003MNRAS.343..459H}
{Hirata} C.,  {Seljak} U.,  2003, \mn@doi [MNRAS]
  {10.1046/j.1365-8711.2003.06683.x}, \href
  {http://adsabs.harvard.edu/abs/2003MNRAS.343..459H} {343, 459}

\bibitem[\protect\citeauthoryear{{Hoekstra} \& {Jain}}{{Hoekstra} \&
  {Jain}}{2008}]{2008ARNPS..58...99H}
{Hoekstra} H.,  {Jain} B.,  2008, \mn@doi [Annual Review of Nuclear and
  Particle Science] {10.1146/annurev.nucl.58.110707.171151}, \href
  {http://adsabs.harvard.edu/abs/2008ARNPS..58...99H} {58, 99}

\bibitem[\protect\citeauthoryear{{Hoekstra}, {Herbonnet}, {Muzzin}, {Babul},
  {Mahdavi}, {Viola}  \& {Cacciato}}{{Hoekstra}
  et~al.}{2015}]{2015MNRAS.449..685H}
{Hoekstra} H.,  {Herbonnet} R.,  {Muzzin} A.,  {Babul} A.,  {Mahdavi} A.,
  {Viola} M.,   {Cacciato} M.,  2015, \mn@doi [\mnras] {10.1093/mnras/stv275},
  \href {http://adsabs.harvard.edu/abs/2015MNRAS.449..685H} {449, 685}

\bibitem[\protect\citeauthoryear{{Hu}}{{Hu}}{2002}]{2002PhRvD..66h3515H}
{Hu} W.,  2002, \mn@doi [\prd] {10.1103/PhysRevD.66.083515}, \href
  {http://adsabs.harvard.edu/abs/2002PhRvD..66h3515H} {66, 083515}

\bibitem[\protect\citeauthoryear{{Hu} \& {Kravtsov}}{{Hu} \&
  {Kravtsov}}{2003}]{hukravtsov03}
{Hu} W.,  {Kravtsov} A.~V.,  2003, \mn@doi [\apj] {10.1086/345846}, \href
  {http://adsabs.harvard.edu/cgi-bin/nph-bib_query?bibcode=2003ApJ...584..702H&db_key=AST}
  {584, 702}

\bibitem[\protect\citeauthoryear{{Huterer}}{{Huterer}}{2002}]{2002PhRvD..65f3001H}
{Huterer} D.,  2002, \mn@doi [\prd] {10.1103/PhysRevD.65.063001}, \href
  {http://adsabs.harvard.edu/abs/2002PhRvD..65f3001H} {65, 063001}

\bibitem[\protect\citeauthoryear{{Johnston} et~al.,}{{Johnston}
  et~al.}{2007a}]{2007arXiv0709.1159J}
{Johnston} D.~E.,  et~al., 2007a, preprint, \href
  {http://adsabs.harvard.edu/abs/2007arXiv0709.1159J} {} (\mn@eprint {arXiv}
  {0709.1159})

\bibitem[\protect\citeauthoryear{{Johnston}, {Sheldon}, {Tasitsiomi},
  {Frieman}, {Wechsler}  \& {McKay}}{{Johnston}
  et~al.}{2007b}]{2007ApJ...656...27J}
{Johnston} D.~E.,  {Sheldon} E.~S.,  {Tasitsiomi} A.,  {Frieman} J.~A.,
  {Wechsler} R.~H.,   {McKay} T.~A.,  2007b, \mn@doi [\apj] {10.1086/510060},
  \href {http://adsabs.harvard.edu/abs/2007ApJ...656...27J} {656, 27}

\bibitem[\protect\citeauthoryear{{Kilbinger} et~al.,}{{Kilbinger}
  et~al.}{2013}]{2013MNRAS.430.2200K}
{Kilbinger} M.,  et~al., 2013, \mn@doi [\mnras] {10.1093/mnras/stt041}, \href
  {http://adsabs.harvard.edu/abs/2013MNRAS.430.2200K} {430, 2200}

\bibitem[\protect\citeauthoryear{{Leauthaud} et~al.,}{{Leauthaud}
  et~al.}{2010}]{2010ApJ...709...97L}
{Leauthaud} A.,  et~al., 2010, \mn@doi [\apj] {10.1088/0004-637X/709/1/97},
  \href {http://adsabs.harvard.edu/abs/2010ApJ...709...97L} {709, 97}

\bibitem[\protect\citeauthoryear{{Li} et~al.,}{{Li} et~al.}{2016}]{lietal15}
{Li} R.,  et~al., 2016, \mn@doi [\mnras] {10.1093/mnras/stw494}, \href
  {http://adsabs.harvard.edu/abs/2016MNRAS.458.2573L} {458, 2573}

\bibitem[\protect\citeauthoryear{{Mandelbaum} et~al.,}{{Mandelbaum}
  et~al.}{2005}]{mandelbaumetal05}
{Mandelbaum} R.,  et~al., 2005, \mn@doi [\mnras]
  {10.1111/j.1365-2966.2005.09282.x}, \href
  {http://adsabs.harvard.edu/abs/2005MNRAS.361.1287M} {361, 1287}

\bibitem[\protect\citeauthoryear{{Mandelbaum}, {Seljak}  \&
  {Hirata}}{{Mandelbaum} et~al.}{2008a}]{2008JCAP...08..006M}
{Mandelbaum} R.,  {Seljak} U.,   {Hirata} C.~M.,  2008a, \mn@doi [\jcap]
  {10.1088/1475-7516/2008/08/006}, \href
  {http://adsabs.harvard.edu/abs/2008JCAP...08..006M} {8, 006}

\bibitem[\protect\citeauthoryear{{Mandelbaum} et~al.,}{{Mandelbaum}
  et~al.}{2008b}]{2008MNRAS.386..781M}
{Mandelbaum} R.,  et~al., 2008b, \mn@doi [\mnras]
  {10.1111/j.1365-2966.2008.12947.x}, \href
  {http://adsabs.harvard.edu/abs/2008MNRAS.386..781M} {386, 781}

\bibitem[\protect\citeauthoryear{{Mandelbaum}, {Seljak}, {Baldauf}  \&
  {Smith}}{{Mandelbaum} et~al.}{2010}]{2010MNRAS.405.2078M}
{Mandelbaum} R.,  {Seljak} U.,  {Baldauf} T.,   {Smith} R.~E.,  2010, \mn@doi
  [\mnras] {10.1111/j.1365-2966.2010.16619.x}, \href
  {http://adsabs.harvard.edu/abs/2010MNRAS.405.2078M} {405, 2078}

\bibitem[\protect\citeauthoryear{{Mandelbaum}, {Hirata}, {Leauthaud}, {Massey}
  \& {Rhodes}}{{Mandelbaum} et~al.}{2012}]{2012MNRAS.420.1518M}
{Mandelbaum} R.,  {Hirata} C.~M.,  {Leauthaud} A.,  {Massey} R.~J.,   {Rhodes}
  J.,  2012, \mn@doi [\mnras] {10.1111/j.1365-2966.2011.20138.x}, \href
  {http://adsabs.harvard.edu/abs/2012MNRAS.420.1518M} {420, 1518}

\bibitem[\protect\citeauthoryear{{Mandelbaum}, {Slosar}, {Baldauf}, {Seljak},
  {Hirata}, {Nakajima}, {Reyes}  \& {Smith}}{{Mandelbaum}
  et~al.}{2013}]{2013MNRAS.432.1544M}
{Mandelbaum} R.,  {Slosar} A.,  {Baldauf} T.,  {Seljak} U.,  {Hirata} C.~M.,
  {Nakajima} R.,  {Reyes} R.,   {Smith} R.~E.,  2013, \mn@doi [\mnras]
  {10.1093/mnras/stt572}, \href
  {http://adsabs.harvard.edu/abs/2013MNRAS.432.1544M} {432, 1544}

\bibitem[\protect\citeauthoryear{{Massey}, {Kitching}  \& {Richard}}{{Massey}
  et~al.}{2010}]{2010RPPh...73h6901M}
{Massey} R.,  {Kitching} T.,   {Richard} J.,  2010, \mn@doi [Reports on
  Progress in Physics] {10.1088/0034-4885/73/8/086901}, \href
  {http://adsabs.harvard.edu/abs/2010RPPh...73h6901M} {73, 086901}

\bibitem[\protect\citeauthoryear{{McKay} et~al.,}{{McKay}
  et~al.}{2001}]{McKayGGL2001}
{McKay} T.~A.,  et~al., 2001, ArXiv Astrophysics e-prints, \href
  {http://adsabs.harvard.edu/abs/2001astro.ph..8013M} {}

\bibitem[\protect\citeauthoryear{{Mehrtens} et~al.,}{{Mehrtens}
  et~al.}{2012}]{2012MNRAS.423.1024M}
{Mehrtens} N.,  et~al., 2012, \mn@doi [\mnras]
  {10.1111/j.1365-2966.2012.20931.x}, \href
  {http://adsabs.harvard.edu/abs/2012MNRAS.423.1024M} {423, 1024}

\bibitem[\protect\citeauthoryear{{Melchior}, {Sutter}, {Sheldon}, {Krause}  \&
  {Wandelt}}{{Melchior} et~al.}{2014}]{MelchiorVoids2014}
{Melchior} P.,  {Sutter} P.~M.,  {Sheldon} E.~S.,  {Krause} E.,   {Wandelt}
  B.~D.,  2014, \mn@doi [\mnras] {10.1093/mnras/stu456}, \href
  {http://adsabs.harvard.edu/abs/2014MNRAS.440.2922M} {440, 2922}

\bibitem[\protect\citeauthoryear{{Miyatake}, {More}, {Takada}, {Spergel},
  {Mandelbaum}, {Rykoff}  \& {Rozo}}{{Miyatake}
  et~al.}{2016}]{2015arXiv150606135M}
{Miyatake} H.,  {More} S.,  {Takada} M.,  {Spergel} D.~N.,  {Mandelbaum} R.,
  {Rykoff} E.~S.,   {Rozo} E.,  2016, \mn@doi [Physical Review Letters]
  {10.1103/PhysRevLett.116.041301}, \href
  {http://adsabs.harvard.edu/abs/2016PhRvL.116d1301M} {116, 041301}

\bibitem[\protect\citeauthoryear{{Nakajima}, {Mandelbaum}, {Seljak}, {Cohn},
  {Reyes}  \& {Cool}}{{Nakajima} et~al.}{2012}]{2012MNRAS.420.3240N}
{Nakajima} R.,  {Mandelbaum} R.,  {Seljak} U.,  {Cohn} J.~D.,  {Reyes} R.,
  {Cool} R.,  2012, \mn@doi [MNRAS] {10.1111/j.1365-2966.2011.20249.x}, \href
  {http://adsabs.harvard.edu/abs/2012MNRAS.420.3240N} {420, 3240}

\bibitem[\protect\citeauthoryear{{Navarro}, {Frenk}  \& {White}}{{Navarro}
  et~al.}{1996}]{nfw96}
{Navarro} J.~F.,  {Frenk} C.~S.,   {White} S.~D.~M.,  1996, \mn@doi [\apj]
  {10.1086/177173}, \href {http://adsabs.harvard.edu/abs/1996ApJ...462..563N}
  {462, 563}

\bibitem[\protect\citeauthoryear{{Navarro}, {Frenk}  \& {White}}{{Navarro}
  et~al.}{1997}]{1997ApJ...490..493N}
{Navarro} J.~F.,  {Frenk} C.~S.,   {White} S.~D.~M.,  1997, \apj, \href
  {http://adsabs.harvard.edu/abs/1997ApJ...490..493N} {490, 493}

\bibitem[\protect\citeauthoryear{{Noh} \& {Cohn}}{{Noh} \&
  {Cohn}}{2012}]{nohcohn12}
{Noh} Y.,  {Cohn} J.~D.,  2012, \mn@doi [\mnras]
  {10.1111/j.1365-2966.2012.21810.x}, \href
  {http://adsabs.harvard.edu/abs/2012MNRAS.426.1829N} {426, 1829}

\bibitem[\protect\citeauthoryear{{Okabe} \& {Smith}}{{Okabe} \&
  {Smith}}{2015}]{okabeetal15}
{Okabe} N.,  {Smith} G.~P.,  2015, preprint, \href
  {http://adsabs.harvard.edu/abs/2015arXiv150704493O} {} (\mn@eprint {arXiv}
  {1507.04493})

\bibitem[\protect\citeauthoryear{{Piccinotti}, {Mushotzky}, {Boldt}, {Holt},
  {Marshall}, {Serlemitsos}  \& {Shafer}}{{Piccinotti}
  et~al.}{1982}]{1982ApJ...253..485P}
{Piccinotti} G.,  {Mushotzky} R.~F.,  {Boldt} E.~A.,  {Holt} S.~S.,  {Marshall}
  F.~E.,  {Serlemitsos} P.~J.,   {Shafer} R.~A.,  1982, \mn@doi [\apj]
  {10.1086/159651}, \href {http://adsabs.harvard.edu/abs/1982ApJ...253..485P}
  {253, 485}

\bibitem[\protect\citeauthoryear{{Planck Collaboration} et~al.,}{{Planck
  Collaboration} et~al.}{2014}]{2014A&A...571A..29P}
{Planck Collaboration} et~al., 2014, \mn@doi [\aap]
  {10.1051/0004-6361/201321523}, \href
  {http://adsabs.harvard.edu/abs/2014A%26A...571A..29P} {571, A29}

\bibitem[\protect\citeauthoryear{{Refregier}}{{Refregier}}{2003}]{2003ARAA..41..645R}
{Refregier} A.,  2003, \mn@doi [\araa]
  {10.1146/annurev.astro.41.111302.102207}, \href
  {http://adsabs.harvard.edu/abs/2003ARA%26A..41..645R} {41, 645}

\bibitem[\protect\citeauthoryear{{Reyes}, {Mandelbaum}, {Gunn}, {Nakajima},
  {Seljak}  \& {Hirata}}{{Reyes} et~al.}{2012}]{2012MNRAS.425.2610R}
{Reyes} R.,  {Mandelbaum} R.,  {Gunn} J.~E.,  {Nakajima} R.,  {Seljak} U.,
  {Hirata} C.~M.,  2012, \mn@doi [MNRAS] {10.1111/j.1365-2966.2012.21472.x},
  \href {http://adsabs.harvard.edu/abs/2012MNRAS.425.2610R} {425, 2610}

\bibitem[\protect\citeauthoryear{{Rowe} et~al.,}{{Rowe}
  et~al.}{2015}]{2015A&C....10..121R}
{Rowe} B.~T.~P.,  et~al., 2015, \mn@doi [Astronomy and Computing]
  {10.1016/j.ascom.2015.02.002}, \href
  {http://adsabs.harvard.edu/abs/2015A%26C....10..121R} {10, 121}

\bibitem[\protect\citeauthoryear{{Rozo} \& {Rykoff}}{{Rozo} \&
  {Rykoff}}{2014}]{rozorykoff14}
{Rozo} E.,  {Rykoff} E.~S.,  2014, \mn@doi [\apj] {10.1088/0004-637X/783/2/80},
  \href {http://adsabs.harvard.edu/abs/2014ApJ...783...80R} {783, 80}

\bibitem[\protect\citeauthoryear{{Rozo} et~al.}{{Rozo}
  et~al.}{2009a}]{rozoetal09a}
{Rozo} E.,  et~al., 2009a, \mn@doi [\apj] {10.1088/0004-637X/699/1/768}, \href
  {http://adsabs.harvard.edu/abs/2009ApJ...699..768R} {699, 768}

\bibitem[\protect\citeauthoryear{{Rozo} et~al.}{{Rozo}
  et~al.}{2009b}]{rozoetal09b}
{Rozo} E.,  et~al., 2009b, \mn@doi [\apj] {10.1088/0004-637X/703/1/601}, \href
  {http://adsabs.harvard.edu/abs/2009ApJ...703..601R} {703, 601}

\bibitem[\protect\citeauthoryear{{Rozo}, {Wu}  \& {Schmidt}}{{Rozo}
  et~al.}{2011}]{rozoetal11}
{Rozo} E.,  {Wu} H.-Y.,   {Schmidt} F.,  2011, \mn@doi [\apj]
  {10.1088/0004-637X/735/2/118}, \href
  {http://adsabs.harvard.edu/abs/2011ApJ...735..118R} {735, 118}

\bibitem[\protect\citeauthoryear{{Rozo} et~al.}{{Rozo}
  et~al.}{2015a}]{rozoetal15}
{Rozo} E.,  et~al., 2015a, \mn@doi [\mnras] {10.1093/mnras/stv605}, \href
  {http://adsabs.harvard.edu/abs/2015MNRAS.450..592R} {450, 592}

\bibitem[\protect\citeauthoryear{{Rozo}, {Rykoff}, {Becker}, {Reddick}  \&
  {Wechsler}}{{Rozo} et~al.}{2015b}]{rozoetal15b}
{Rozo} E.,  {Rykoff} E.~S.,  {Becker} M.,  {Reddick} R.~M.,   {Wechsler} R.~H.,
   2015b, \mn@doi [\mnras] {10.1093/mnras/stv1560}, \href
  {http://adsabs.harvard.edu/abs/2015MNRAS.453...38R} {453, 38}

\bibitem[\protect\citeauthoryear{{Rykoff} et~al.}{{Rykoff}
  et~al.}{2012}]{rykoffetal12}
{Rykoff} E.~S.,  et~al., 2012, \mn@doi [\apj] {10.1088/0004-637X/746/2/178},
  \href {http://adsabs.harvard.edu/abs/2012ApJ...746..178R} {746, 178}

\bibitem[\protect\citeauthoryear{{Rykoff} et~al.,}{{Rykoff} et~al.}{2014}]{rmI}
{Rykoff} E.~S.,  et~al., 2014, \mn@doi [\apj] {10.1088/0004-637X/785/2/104},
  \href {http://adsabs.harvard.edu/abs/2014ApJ...785..104R} {785, 104}

\bibitem[\protect\citeauthoryear{{Rykoff}, {Rozo}  \& {Keisler}}{{Rykoff}
  et~al.}{2015}]{2015arXiv150900870R}
{Rykoff} E.~S.,  {Rozo} E.,   {Keisler} R.,  2015, preprint, \href
  {http://adsabs.harvard.edu/abs/2015arXiv150900870R} {} (\mn@eprint {arXiv}
  {1509.00870})

\bibitem[\protect\citeauthoryear{{Rykoff} et~al.,}{{Rykoff}
  et~al.}{2016}]{2016arXiv160100621R}
{Rykoff} E.~S.,  et~al., 2016, \mn@doi [\apjs] {10.3847/0067-0049/224/1/1},
  \href {http://adsabs.harvard.edu/abs/2016ApJS..224....1R} {224, 1}

\bibitem[\protect\citeauthoryear{{Saro} et~al.,}{{Saro}
  et~al.}{2015}]{2015MNRAS.454.2305S}
{Saro} A.,  et~al., 2015, \mn@doi [\mnras] {10.1093/mnras/stv2141}, \href
  {http://adsabs.harvard.edu/abs/2015MNRAS.454.2305S} {454, 2305}

\bibitem[\protect\citeauthoryear{{Schaller} et~al.,}{{Schaller}
  et~al.}{2015}]{schaller15}
{Schaller} M.,  et~al., 2015, \mn@doi [\mnras] {10.1093/mnras/stv1067}, \href
  {http://adsabs.harvard.edu/abs/2015MNRAS.451.1247S} {451, 1247}

\bibitem[\protect\citeauthoryear{{Schmidt}, {Rozo}, {Dodelson}, {Hui}  \&
  {Sheldon}}{{Schmidt} et~al.}{2009}]{2009PhRvL.103e1301S}
{Schmidt} F.,  {Rozo} E.,  {Dodelson} S.,  {Hui} L.,   {Sheldon} E.,  2009,
  \mn@doi [Physical Review Letters] {10.1103/PhysRevLett.103.051301}, \href
  {http://adsabs.harvard.edu/abs/2009PhRvL.103e1301S} {103, 051301}

\bibitem[\protect\citeauthoryear{{Schneider}}{{Schneider}}{2006}]{2006glsw.conf....1S}
{Schneider} P.,  2006, in {Meylan} G.,  {Jetzer} P.,  {North} P.,  {Schneider}
  P.,  {Kochanek} C.~S.,   {Wambsganss} J.,  eds, Saas-Fee Advanced Course 33:
  Gravitational Lensing: Strong, Weak and Micro. pp 1--89

\bibitem[\protect\citeauthoryear{{Sereno}, {Fedeli}  \& {Moscardini}}{{Sereno}
  et~al.}{2016}]{2016JCAP...01..042S}
{Sereno} M.,  {Fedeli} C.,   {Moscardini} L.,  2016, \mn@doi [\jcap]
  {10.1088/1475-7516/2016/01/042}, \href
  {http://adsabs.harvard.edu/abs/2016JCAP...01..042S} {1, 042}

\bibitem[\protect\citeauthoryear{{Sheldon} et~al.,}{{Sheldon}
  et~al.}{2004a}]{2004AJ....127.2544S}
{Sheldon} E.~S.,  et~al., 2004a, \mn@doi [\aj] {10.1086/383293}, \href
  {http://adsabs.harvard.edu/abs/2004AJ....127.2544S} {127, 2544}

\bibitem[\protect\citeauthoryear{{Sheldon} et~al.,}{{Sheldon}
  et~al.}{2004b}]{SheldonGGL2004}
{Sheldon} E.~S.,  et~al., 2004b, \mn@doi [\aj] {10.1086/383293}, \href
  {http://adsabs.harvard.edu/abs/2004AJ....127.2544S} {127, 2544}

\bibitem[\protect\citeauthoryear{{Sheldon}, {Cunha}, {Mandelbaum}, {Brinkmann}
  \& {Weaver}}{{Sheldon} et~al.}{2012}]{SheldonPhotoz2012}
{Sheldon} E.~S.,  {Cunha} C.~E.,  {Mandelbaum} R.,  {Brinkmann} J.,   {Weaver}
  B.~A.,  2012, \mn@doi [\apjs] {10.1088/0067-0049/201/2/32}, \href
  {http://adsabs.harvard.edu/abs/2012ApJS..201...32S} {201, 32}

\bibitem[\protect\citeauthoryear{{Sif{\'o}n}, {Hoekstra}, {Cacciato}, {Viola},
  {K{\"o}hlinger}, {van der Burg}, {Sand}  \& {Graham}}{{Sif{\'o}n}
  et~al.}{2015}]{2015A&A...575A..48S}
{Sif{\'o}n} C.,  {Hoekstra} H.,  {Cacciato} M.,  {Viola} M.,  {K{\"o}hlinger}
  F.,  {van der Burg} R.~F.~J.,  {Sand} D.~J.,   {Graham} M.~L.,  2015, \mn@doi
  [\aap] {10.1051/0004-6361/201424435}, \href
  {http://adsabs.harvard.edu/abs/2015A%26A...575A..48S} {575, A48}

\bibitem[\protect\citeauthoryear{{Simet} \& {Mandelbaum}}{{Simet} \&
  {Mandelbaum}}{2015}]{2015MNRAS.449.1259S}
{Simet} M.,  {Mandelbaum} R.,  2015, \mn@doi [\mnras] {10.1093/mnras/stv313},
  \href {http://adsabs.harvard.edu/abs/2015MNRAS.449.1259S} {449, 1259}

\bibitem[\protect\citeauthoryear{{Singh}, {Mandelbaum}  \& {More}}{{Singh}
  et~al.}{2015}]{2015MNRAS.450.2195S}
{Singh} S.,  {Mandelbaum} R.,   {More} S.,  2015, \mn@doi [\mnras]
  {10.1093/mnras/stv778}, \href
  {http://adsabs.harvard.edu/abs/2015MNRAS.450.2195S} {450, 2195}

\bibitem[\protect\citeauthoryear{{Springel}}{{Springel}}{2005}]{Springel2005}
{Springel} V.,  2005, \mn@doi [\mnras] {10.1111/j.1365-2966.2005.09655.x},
  \href {http://adsabs.harvard.edu/abs/2005MNRAS.364.1105S} {364, 1105}

\bibitem[\protect\citeauthoryear{{Wechsler}, {Zentner}, {Bullock}, {Kravtsov}
  \& {Allgood}}{{Wechsler} et~al.}{2006}]{2006ApJ...652...71W}
{Wechsler} R.~H.,  {Zentner} A.~R.,  {Bullock} J.~S.,  {Kravtsov} A.~V.,
  {Allgood} B.,  2006, \mn@doi [\apj] {10.1086/507120}, \href
  {http://adsabs.harvard.edu/abs/2006ApJ...652...71W} {652, 71}

\bibitem[\protect\citeauthoryear{{Wright} \& {Brainerd}}{{Wright} \&
  {Brainerd}}{2000}]{WrightBrainerd}
{Wright} C.~O.,  {Brainerd} T.~G.,  2000, \mn@doi [\apj] {10.1086/308744},
  \href {http://adsabs.harvard.edu/abs/2000ApJ...534...34W} {534, 34}

\bibitem[\protect\citeauthoryear{{Yang}, {Mo}, {van den Bosch}, {Jing},
  {Weinmann}  \& {Meneghetti}}{{Yang} et~al.}{2006}]{2006MNRAS.373.1159Y}
{Yang} X.,  {Mo} H.~J.,  {van den Bosch} F.~C.,  {Jing} Y.~P.,  {Weinmann}
  S.~M.,   {Meneghetti} M.,  2006, \mn@doi [\mnras]
  {10.1111/j.1365-2966.2006.11091.x}, \href
  {http://adsabs.harvard.edu/abs/2006MNRAS.373.1159Y} {373, 1159}

\bibitem[\protect\citeauthoryear{{York} et~al.,}{{York}
  et~al.}{2000}]{2000AJ....120.1579Y}
{York} D.~G.,  et~al., 2000, \mn@doi [\aj] {10.1086/301513}, \href
  {http://adsabs.harvard.edu/abs/2000AJ....120.1579Y} {120, 1579}

\bibitem[\protect\citeauthoryear{{van Uitert}, {Gilbank}, {Hoekstra},
  {Semboloni}, {Gladders}  \& {Yee}}{{van Uitert}
  et~al.}{2016}]{2016A&A...586A..43V}
{van Uitert} E.,  {Gilbank} D.~G.,  {Hoekstra} H.,  {Semboloni} E.,  {Gladders}
  M.~D.,   {Yee} H.~K.~C.,  2016, \mn@doi [\aap] {10.1051/0004-6361/201526719},
  \href {http://adsabs.harvard.edu/abs/2016A%26A...586A..43V} {586, A43}

\bibitem[\protect\citeauthoryear{{von der Linden}, {Best}, {Kauffmann}  \&
  {White}}{{von der Linden} et~al.}{2007}]{2007MNRAS.379..867V}
{von der Linden} A.,  {Best} P.~N.,  {Kauffmann} G.,   {White} S.~D.~M.,  2007,
  \mn@doi [\mnras] {10.1111/j.1365-2966.2007.11940.x}, \href
  {http://adsabs.harvard.edu/abs/2007MNRAS.379..867V} {379, 867}

\bibitem[\protect\citeauthoryear{{von der Linden} et~al.,}{{von der Linden}
  et~al.}{2014}]{2014MNRAS.439....2V}
{von der Linden} A.,  et~al., 2014, \mn@doi [\mnras] {10.1093/mnras/stt1945},
  \href {http://adsabs.harvard.edu/abs/2014MNRAS.439....2V} {439, 2}

\makeatother
\end{thebibliography}

\appendix
\section{Details of $\Delta\Sigma$ model components}\label{app:models}

Our final model includes marginalization over multiple modeling issues that are secondary to the
relationship we are primarily interested in measuring.  We show here a simple model including none of
them, and then add more complications one at a time, to better show the effect each additional parameter has
on our final results.  Table~\ref{mcresults-science} shows the results 
for our MCMC model fitting for different combinations of parameters.  We discuss here the various complications of the model.

\begin{table*}
\centering
\begin{tabular}{l | c c c | c c}
	 & No. of steps & Minimum & & & \\
Model name & (burn-in)  &
$-2\ln \lkhd$ & d.o.f. & $\log_{10} M_0^*$ & $\alpha$ \\
\hline
\footnotesize{Simple}  & 1000 (100) & 78.41 & 75 & $14.322^{+0.028}_{-0.029}$ & $1.36^{+0.10}_{-0.09}$ \\
\footnotesize{Simple with $M-c$ scatter}  & 1000 (100) & 78.09 & 75 & $14.323^{+0.028}_{-0.029}$ & $1.36^{+0.09}_{-0.10}$  \\
\footnotesize{Fitted $M-c$ amplitude \& $M-c$ scatter} & 1500 (100) & 77.64 & 74 & $14.329 \pm 0.030$ & $1.38^{+0.09}_{-0.10}$   \\
\footnotesize{Miscentring \& $M-c$ scatter} & 1500 (100) & 78.08 & 73 & $14.347 \pm 0.030$ & $1.36^{+0.09}_{-0.10}$  \\
\footnotesize{Miscentring, fitted $M-c$ amplitude \& $M-c$ scatter} & 2250 (100) & 76.37 & 72 & $14.344 \pm 0.031$ & $1.33^{+0.09}_{-0.10}$ \\
\end{tabular}
\caption{Fitting results for all of our cluster models, showing the 
parameters of scientific interest only.   We 
report the settings for the MCMC chain and the maximum likelihood $-2\ln \lkhd$ of any step in any chain 
for that MCMC run.  We run the \texttt{emcee} MCMC ensemble sampler with 100 walkers for every chain,
so the total number of steps included in the plots is 100 times what is listed here (including burn-in steps).  
When the Gaussian priors are not important, $-2\ln \lkhd$ is equal to the $\chi^2$.
We also report the median value and the $1\sigma$ error regions (the 16th-84th percentile range) for the 
parameters of scientific interest.
All models are NFW (Eq.~\ref{NFW-rho}), some with miscentring 
(Eq.~\ref{miscentring}), some with fitted $M-c$ relations.  The last model in the table is our fiducial model
used in the analysis in the main text of this work.
The $^*$ symbol indicates that the mass values from the output of the MCMC chain have been corrected for model bias, as described
in Section~\ref{sec:sys}, and the error bar is combined statistical and systematic.}
\label{mcresults-science}
\end{table*}

Our simple model assumes no miscentring and a mass--concentration ($M-c$) relation from the literature,
\citet{2013ApJ...766...32B}, evaluated using the publicly-available \texttt{colossus} package described in 
\citet{2015ApJ...799..108D}, without $M-c$ scatter.  The values we obtain
for the parameters, shown in Table~\ref{mcresults-science}, already have many of the characteristics 
we see in the final analysis: $\log M_0$ is somewhat above 14 and $\alpha$ well above 1.  Our parameter values 
do not change significantly when we add the various complications: lognormal scatter in the 
mass--concentration relation with width 0.14 dex as described in the main text; a variable mass--concentration relation;
miscentring; and all three.  There is a $\sim 0.3\sigma$ decrease in $\alpha$ when
we include all effects versus all other models, indicating that a small amount of the apparent scaling of mass with 
$\lambda$ in the simple model can
be attributed to those effects changing with the cluster richness, but the effect is small.  The nuisance parameters
($b$ and miscentring) are largely determined by our priors; $c_0$, the amplitude of the 
mass--concentration relation, shows more movement, with a value of $0.90^{+0.12}_{-0.10}$ when there is
no miscentring included but $1.34^{+0.35}_{-0.25}$ when it is.  Adequate modeling of the miscentring is 
clearly important for weak lensing studies such as these which attempt to measure the $M-c$ relation using 
information at scales less than the miscentring radius.

The fit quality does not vary strongly with the inclusion of more
parameters, as judged by the best-fit likelihood.  In summary, we can
fit the \redmapper{} cluster lensing signal reasonably well with a sum
of NFW profiles that have characteristics determined by the input
$\lambda$ and redshift distribution of the \redmapper{} catalogue.
For the radius range and cluster distribution included in this work,
including higher-level complications such as miscentring and variable
mass--concentration relations are not important to obtaining a
reliable mass amplitude, but do affect the scaling with richness to
some extent.

\section{Redshift evolution}\label{app:redshift}

To test for possible redshift evolution of the mass--richness relation across our redshift range, instead of Eq.~\eqref{massmodel}, we fit a model of the form
\begin{equation}\label{eq:zmodel}
M = M_0 \left(\frac{\lambda}{\lambda_0}\right)^{\alpha}\left(\frac{1+z}{1+z_0}\right)^{\beta}
\end{equation}
with $z_0=0.2$.  We then perform an MCMC run of the simple model from Appendix~\ref{app:models} (with
only $M_0$, $\alpha$, scatter, and $b$), including this $\beta$ parameter as one of the variables we want to fit, 
with a top-hat prior of $(-12,2)$; the prior is asymmetric because early testing in the range $(-1,1)$ did not reveal an
obvious peak but a preference for values $\leq 1$, and we were unsure how low we would need to go to ensure it was included. 

The value preferred for $\beta$, $-1.5 \pm 0.9$, is consistent with
our expected value of 0 at slightly above $1.5\sigma$.  The fit is
better, with the minimum $-2 \ln \lkhd=76.64$ instead of 78.41, but no
other parameter moves significantly when $\beta$ is included.  Since
we expect $\beta \approx 0$ and cannot exclude this possibility using
the data, we do not include redshift evolution of the mass--$\lambda$
relation in the model used for our main results in this paper.  Fixing
$\beta=0$ is a stronger assumption than letting $\beta$ float with
some prior on expected values; however, we know what fixing $\beta=0$
does (finds the average value over all redshifts) while it is unclear
what we would be modeling if we allowed $\beta$ to float, so we make
the tradeoff of being insensitive to redshift in favor of better
understanding the physical importance of our other parameters.

\section{The Projection Rate in the \redmapper\ Cluster Catalogue}
\label{app:proj}

In section \ref{sec:proj}, we estimated the impact of projection
effects on our weak lensing mass calibration.  Our starting point in
that section was the projection rate in the \redmapper\ cluster
catalogue.  Here, we discuss how we arrived at the prior for this
projection rate.

Our starting point is \citet{rmI}, who set a lower limit on the
incidence of these type of projection effects by placing synthetic
galaxy clusters within the SDSS survey and then performing a cluster
finding step.  We found a richness-dependent rate of projection
effects that increases with richness, reaching 5\% for the richest
clusters.

In practice, this fraction should be boosted due to correlated structures along the line of sight.
The associated multiplicative factor $\eta$ should take the form
\begin{equation}
\eta = \frac{\int \mathrm{d}^3V\ (1+ \xi)}{\int \mathrm{d}^3 V}
\end{equation}
where $\xi$ is the halo-halo correlation function, and the integrals are over the volume over which projection effects can
occur.  We model this volume as a cylinder of radius $R_{\rm{min}}\approx 2\ \Mpc$ and a height $2\sigma_z$,
where $\sigma_z$ is the photometric redshift uncertainty of red galaxies.  This corresponds to a cylinder
height $R_{\rm{max}} \approx 130\ \Mpc$.  Adopting a correlation function $\xi \propto r^{-2}$, and 
a typical cluster correlation length $R_{\rm corr} \approx 15\ \Mpc$ 
we arrive at
\begin{equation}
\eta = 1 + \frac{R_{\rm corr}^2}{R_{\rm{min}}R_{\rm{max}}} \approx 1.8.
\end{equation}
Adopting a $\sim 50\%$ systematic uncertainty for the clustering correction,
so that $\eta=1.8 \pm 0.4$, 
we arrive at a total projection rate of $\eta\times 5\% = 9\% \pm 2\%$ for the highest richness
clusters.

A second estimate can be made on the basis of \citetalias{rozoetal15b}, who find that $6\% \pm 2\%$ of the
cluster richness is due to projected galaxies in excess of the background expectation.  We assuming these
galaxies are due to projections of two comparably rich systems, each contributing $\approx \lambda/2$
galaxies to a single blended cluster of richness $\lambda$.  In this case, the fraction of projected
galaxies $f=0.06 \pm 0.02$ is given by
\begin{equation}
f = \frac{pn(\lambda)(\lambda/2)}{n(\lambda)\lambda} = p/2.
\end{equation}
where $n(\lambda)$ is the abundance of clusters of richness $\lambda$.  We arrive then
at $p= 12\% \pm 4\%$, in reasonable agreement with our original estimate.  
A simulation analysis by Kitadinis et al. (in preparation) find a similar projection rate for
\redmapper\ clusters in simulations.  

On the basis of these arguments, we adopt a projection rate $p=10\% \pm 4\%$.

\end{document}